\documentclass[twocolumn, prb]{revtex4}

\usepackage{graphicx}
\usepackage{dcolumn}
\usepackage{bm}

\begin{document}

\title{ A two-step density-matrix renormalization-group study of coupled Luttinger liquids} 

\author { S. Moukouri, E. Eidelstein}

\affiliation{ Racah Institute of Physics, Hebrew University, Jerusalem 91904 Israel} 

\begin{abstract}
We report a two-step density-matrix renormalization-group computation of the equal-time single-particle Green's function, 
 the density-density correlations, and the low-frequency spectral weight function of a spinless fermion model 
in an anisotropic two-dimensional lattice
at half-filling. We find that at weak couplings the density-density correlations have the universal decay 
of a Fermi liquid; the spectral weight function displays a sharp quasi-particle peak. But in the
vicinity of a quantum critical point,  these correlations strongly deviate
from a Fermi liquid prediction and a pseudogap opens in the  spectral weight function.
\end{abstract}

\maketitle

\section{Introduction}

The metallic phase of interacting three-dimensional (3D) electron systems is described by
 the Fermi liquid theory (FLT) proposed by L. Landau \cite{landau}. The Fermi
liquid (FL) is a phase 
of matter  in which low energy charged excitations and the long-distance behavior of correlation functions
are essentially similar to those of the non-interacting electron system. They carry spin and are described 
as weakly-interacting quasi-particles. 
 A microscopic justification of the FLT 
was given through the many-body perturbation theory \cite{agd,nozieres} and
the renormalization group (RG) \cite{shankar,chitov1,chitov2}. However neither the many-body
perturbation theory nor the RG provide a complete demonstration of the
emergence of a FL from an interacting electron model. The many-body perturbation theory neglects the competition between different channels and focuses only
on the metallic phase. The RG analyses the flow of the interaction toward
the FL fixed point, but it does not yield the quasi-particle spectra.
Furthermore, both approaches are restricted to the weak-coupling regime.
Indisputable FL behavior has been theoretically shown only
for impurity models \cite{nozieres2} or model in infinite 
dimensions \cite{kotliar}. Recent development into this difficult problem
involved the use of string theory \cite{zaanen}.

Electron-electron interactions have a dramatic effect in the one-dimensional (1D) metallic phase. No matter how small, they completely
destroy the quasi-particles. An alternative to the FLT for 1D metals is Haldane's \cite{haldane}  Luttinger
liquid theory (LLT). The central assumption of the LLT is that the low-energy excitations and the long-distance behavior of correlation
functions of 1D metals are similar
to those of a model introduced by Luttinger \cite{luttinger}. These excitations 
 are density fluctuations which propagate with different velocities for the spin and charge. In a Luttinger liquid (LL), unlike a FL, the decay of the correlation functions is non-universal.

There is a significant interest in the question of the evolution of the Luttinger 
liquid when going from $D=1$ to $D>1$. This is relevant to the physics of 
quasi-one-dimensional organic conductors \cite{bourbonnais} 
 where pressure or temparuture can induce a crossover from an LL to a FL or an
ordered phase. The dimensional crossover has been studied by various approaches. These include analytic continuation 
from D=1 to $D=1+\epsilon$ \cite{castellani}, perturbative renormalization group (RG) on weakly-coupled LL\cite{bourbonnais}
or on 2D system with weak interaction \cite{shankar},  functional 
integral \cite{boies}, and generalized dynamical mean-field theory (DMFT) \cite{arrigoni,bierman}. 
These studies conclude to a FL ground state in 2D. 
 However, Anderson and coworkers \cite{clarke,strong} have argued that, 
a different scenario due to strong interaction could take place. They
pointed out that despite the RG being relevant, the resulting 2D system
could nevertheless be a non-FL. The effects of the interactions could be
so dramatic that if the transverse hopping is not strong enough, the
electrons would remain confined in the chains. Coherent quasi-particles
would form only when the transverse hopping exceeds a treshold. This
issue has recently been reexaminated in the framework of the functional
RG \cite{ledowski}. A regime with confined coherence was predicted  in 
the strong interaction regime. 
The non-FL mechanism suggested in Ref.\cite{clarke,strong} could occur 
for instance in the vicinity 
of a quantum critical point (QCP) where interaction effects are very
strong.  
 
  In this paper, we use the two-step density-matrix renormalization
group method \cite{moukouri-TSDMRG} to study the possible emergence of FL
and non-FL behaviors on an interacting electron model close to a QCP. 
Our results are 
consistent with a FL 
ground state in a 2D model for weak interactions. We also show that as a QCP is
approached,  the system enters a non-FL regime. This is captured by the behavior of the 
exponent $K$ of the density-density correlation which shows a strong
renormalization towards its FL value for $V \alt 1$ and which is
only weakly renormalized in the vicinity of the 1D quantum critical
point.
The evolution from a FL to a non-FL is also oberved in the low frequency
spectral weight function.
 
\section{Model and Method}

We concentrate on the following quasi-one-dimensional spinless fermion model on a finite lattice
of size $L_x$, $L_y$ in the $x$, $y$ directions respectively:

\begin{eqnarray}
\nonumber H=-t_x\sum_{i_x,i_y}(c_{i_x,i_y}^{\dagger}c_{i_x+1,i_y}+h.c.)-\mu \sum_{i_x,i_y} n_{i_x,i_y}-\\
t_y\sum_{i_x,i_y}(c_{i_x,i_y}^{\dagger}c_{i_x,i_y+1}+h.c.)+V\sum_{i_x,i_y}n_{i_x,i_y} n_{i_x+1,i_y} .  
\label{hamiltonian}
\end{eqnarray}

\noindent We are interested in the situation where the hopping parameter $t_x$ along the $x$ direction 
is far larger than the interchain hopping $t_y$, $t_x \gg t_y$. The interaction $V$
is chosen such that when $t_y=0$, we are in the LL phase, i.e., $V \alt 2t_x$. We will restrict
ourselves to case where the electron density is at half-filling, $n_0=N_e/(L_xL_y)=1/2$, where $N_e$ is the
total number of electrons. It has been shown in Ref.~\onlinecite{moukouri-TSDMRG} that this type of anisotropic model may  be studied
 using the density-matrix renormalization group (DMRG) method \cite{white}. 
In this approach, the DMRG is applied in two steps.

In the first step, we use the 
DMRG to construct an approximate, yet well controlled, low-energy Hamiltonian ${\tilde H}_{0,i_y}$ 
for an isolated chain $i_y$ Hamiltonian 

\begin{eqnarray}
\nonumber H_{0,i_y}=-t_x\sum_{i_x}(c_{i_x,i_y}^{\dagger}c_{i_x+1,i_y}+h.c.)-\mu \sum_{i_x} n_{i_x,i_y}+\\
V\sum_{i_x}n_{i_x,i_y} n_{i_x+1,i_y}.
\end{eqnarray}

\noindent In order to allow interchain dynamics, ${\tilde H}_{0,i_y}$ is obtained by targeting the ground state of 
the nominal filling $N_{ex}/L_x=1/2$, where $N_{ex}$ is the number of electrons on the chain $i_y$. We also target ground states of 
$N_{ex} \pm 1$, $N_{ex} \pm 2$, $N_{ex} \pm 3$,... until the lowest  state of a sector $N_{ex} \pm k$ is
higher than the highest state kept in the $N_{ex}$ sector.

In the second step, the  full 2D 
Hamiltonian (\ref{hamiltonian}) is projected onto the basis constructed from the tensor product of the 
single-chain eigenfunctions; this projection yields an effective one-dimensional Hamiltonian for the 2D lattice,

\begin{eqnarray}
 \tilde{H} \approx \sum_{i_y} \tilde{H}_{0,i_y} -t_y\sum_{i_y}(\tilde{c}_{i_y}^{\dagger} \tilde{c}_{i_y+1}+h.c.).
 \end{eqnarray}
 
 \noindent $\tilde{H}_{0,i_y}$ is diagonal, its element are the DMRG eigenvalues. $\tilde{c}\textbf{\textbf{}}_{i_y}^{\dagger}$, 
$\tilde{c}_{i_y}$, and $\tilde{n}_{i_y}$ are the  renormalized operators in the single chain basis. These 
are vector operators made of local operators on each site of a chain $i_y$. It is clear that during the passage
from the first to the second step, this method is different from the conventional DMRG in that the truncation
is not done through the reduced density matrix. The truncation is done rather like in the real space RG method \cite{weinstein}. 
But if $t_y$ remains small with respect to the energy width of the states kept, as in the Wilson approach for the
Kondo problem \cite{wilson}, this algorithm can retain high accuracy as we will show below.

\begin{table}
\begin{ruledtabular}
\begin{tabular}{cccc}
    & $16 \times 17$ & $32 \times 33$ & $64 \times 65$ \\
\hline
 $\Delta E$ & $1.950841$ & $1.035854$ & $0.534903$  \\
 $\delta \epsilon (t_y=0.05t_x)$ & $2.0 \times 10^{-8}$ & $< 10^{-8}$ & $2.4 \times 10^{-7}$  \\
 $\delta \epsilon (t_y=0.1t_x)$ & $1.0 \times 10^{-8}$ & $4.6 \times 10^{-7}$ & $2.9 \times 10^{-5}$  \\
\end{tabular}
\end{ruledtabular}
\caption{Energy width $\Delta E$ of the $m_2=80$ states kept and error $\delta \epsilon$ in 
the ground-energy per site as function of the lattice size.} 
\label{energydata}
\end{table}

\section{Test on the non-Interacting case}

Let us first analyze the performance of this DMRG algorithm for the case $V=0$ which enjoys
an exact solution. The exact single particle energies and wave functions for open
boundary conditions are respectively:

\begin{eqnarray}
\epsilon_{{\bf k}_{l,m}}= -2t_x cos k_{x_l}-2t_ycos k_{y_m},
\end{eqnarray} 

\begin{eqnarray}
\psi_{{\bf k}_{l,m}}(i_x,i_y)=2 \frac{sin (k_{x_l}i_x) sin(k_{y_m}i_y)}
{\sqrt{(L_x+1)(L_y+1)}}, 
\end{eqnarray}

\noindent where
$k_{x_l}=l\pi/(L_x+1)$, $k_{y_m}=m\pi/(L_y+1)$, $l=1,...,L_x$, $m=1,...,L_y$. 
The ground-state energy

\begin{eqnarray}
E_0=\sum_{|{\bf k}_{l,m}|<k_F} \epsilon_{{\bf k}_{l,m}},
\end{eqnarray} 

\noindent the single particle
Green's function between two points of coordinates $(i_x,i_y)$ and $(j_x,j_y)$, 

\begin{eqnarray}
g_0((i_x,i_y);(j_x,j_y))=<c^{\dagger}_{i_x,i_y}c_{j_x,j_y}>,
\end{eqnarray}

\noindent and the density-density correlation between
these points 

\begin{eqnarray}
C_0((i_x,i_y);(j_x,j_y))=<(n_{i_x,i_y}-n_0)(n_{j_x,j_y}-n_0)>,
\end{eqnarray}

\noindent with 
$n_{\alpha_x,\alpha_y}=c^{\dagger}_{\alpha_x,\alpha_y}c_{\alpha_x,\alpha_y}$, 
may be readily computed. 

\begin{eqnarray}
g_0((i_x,i_y);(j_x,j_y))=\sum_{|{\bf k}_{l,m}|<k_F}\psi_{\bf k}^*(i_x,i_y)
\psi_{\bf k}(j_x,j_y), 
\end{eqnarray}

\noindent $C_0((i_x,i_y);(j_x,j_y))$ is obtained from $g_0((i_x,i_y);(j_x,j_y))$ by the using Wick's theorem.

In comparing the DMRG to this exact result, we emphasize that although the exact solution is trivial in momentum 
representation, for a real space method such as DMRG it remains a difficult challenge. However, unlike the exact solution,
the DMRG can readily be extended to the case $V \neq 0$ without difficulty. In the DMRG, we kept
 up to $m_1=384$ states during the first step with up to $L_x=130$. For this value of $m_1$, the truncation
 error is virtually zero. Among the $4\times m_1^2$ states of the superblock, we kept
 a subset of up to $m_2=108$ states during the second step. These yield size dependent energy widths
 $\Delta E=\epsilon_{m_2}-\epsilon_0$, where $\epsilon_0$ and $\epsilon_{m_2}$
are respectively the lowest state and the highest state kept in a chain $i_y$
 (see Table \ref{energydata}). The key to retain accuracy during the second step
 is to choose $t_y$ such that, $t_y \ll \Delta E$ for a given $L_x$. We show for instance in Table \ref{energydata}
 the error in the ground state energies  for $t_y=0.05 t_x$ and $t_y=0.1t_x$ for 
 $L_x \times L_y=16 \times 17,~32 \times 33,~64\times 65$. For $t_y=0.05 t_x$, there is an excellent agreement with the exact energy
 for all sizes shown. Note that for the $16 \times 17$ and $32 \times 33$ lattices, 
 the limitation to only 8 digits is due to the fact that we set the error to $10^{-6}$ in the diagonalization of 
  the Hamiltonian. We could easily reach a smaller error
 without significant additional work. The agreement remains excellent for $t_y=0.1t_x$ except for a $64 \times 65$ 
 lattice. At this size, the difference between the DMRG and the  exact energies is two orders of magnitude larger. The crossover temperature from 1D to 2D is
given by $T_X \approx t_y/\pi$ \cite{bourbonnais}, for finite size systems in the ground state, this
translates to $\delta E \approx t_y$, where $\delta E$ is the size-dependent
lowest excitation in 1D. We find that if $\delta E/t_y \alt 1$ and 
$\Delta E/t_y \approx 10$, the
accuracy is almost independent of the system size. For instance, when
$L_x \times L_y= 32 \times ~33, 64 \times ~65, 130 \times ~131$,
 we find respectively, $\delta E=0.0951, ~0.0483, ~0.0240$. By
respectively choosing $t_y=0.1,~0.05, ~0.025$, we obtained 
$\delta \epsilon \alt 10^{-6}$. Hence, if we keep the ratios $\delta E/t_y$ and
$\Delta E/t_y$ constant, we can access the 2D regime in
large systems while retaining very good accuracy.

\begin{figure}
\begin{center}
\includegraphics[width=5.cm, height=4.cm]{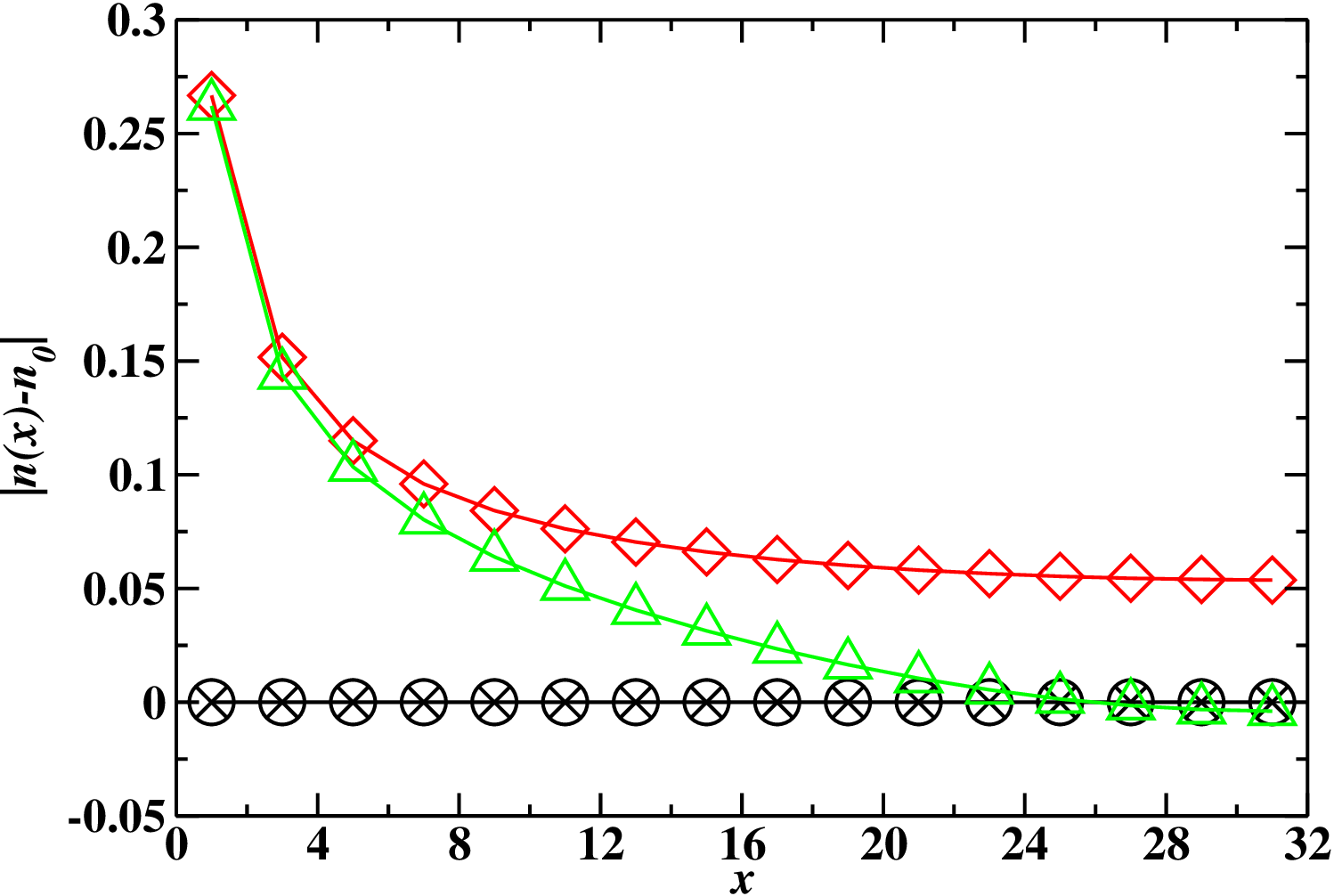}
\end{center}
\caption{
 Density correlation $n(x)$ as function the position $x$ for the middle chain of 2D 
systems: DMRG with $V=0$, $t_y=0.05$ (circles); the exact result with $V=0$, $t_y=0.05$ 
(crosses); interacting 2D system with $V=1.5$, $t_y=0.1$ (triangles); single chain with 
$V=1.5$ (diamonds). }
\label{local}
\end{figure}

\begin{figure}
\begin{center}
$\begin{array}{c@{\hspace{0.5in}}c}
         \multicolumn{1}{l} {}\\ [-0.23cm]
\includegraphics[width=4.cm, height=3.cm]{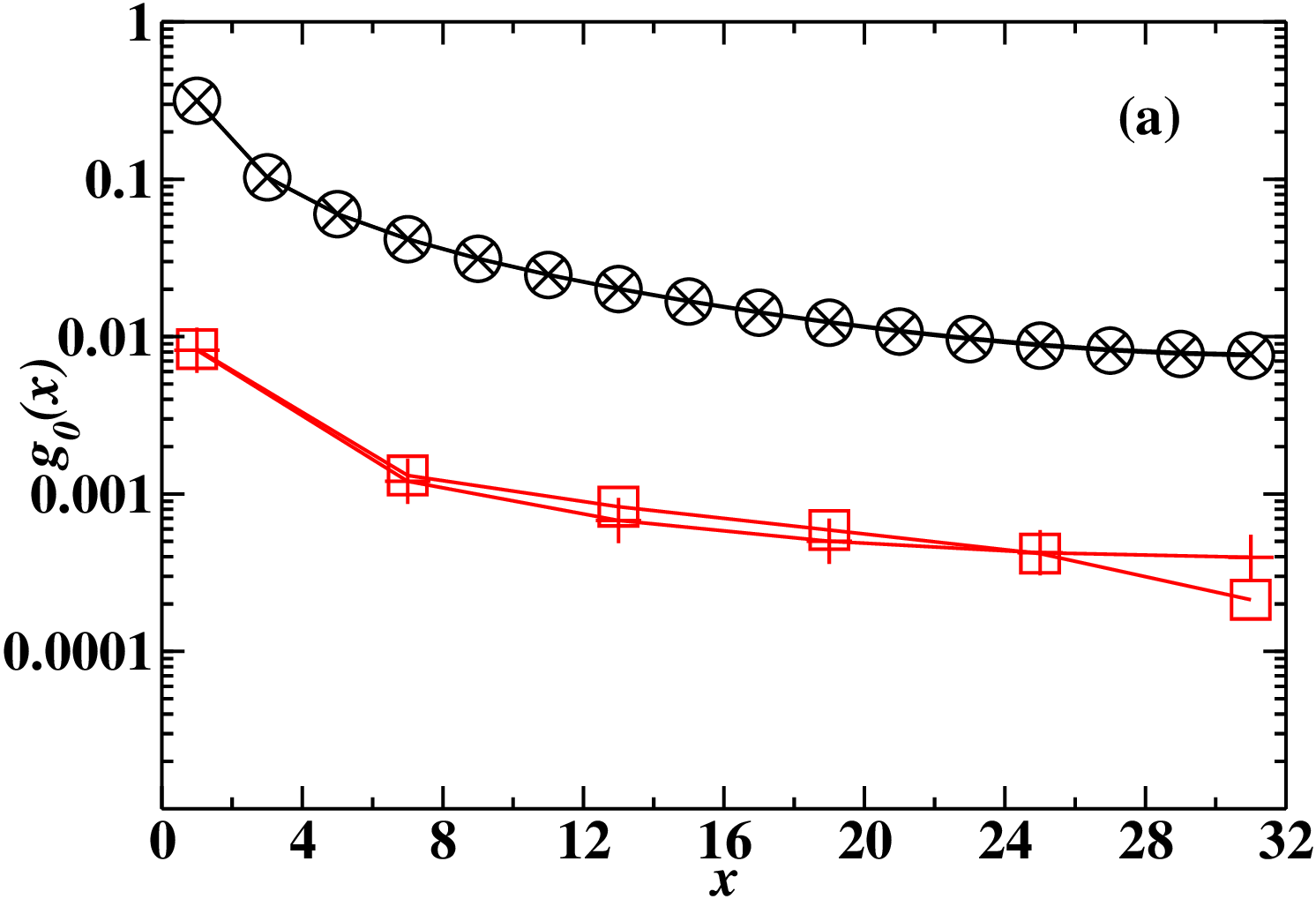}
\hspace{0.5cm}
\vspace{0.5cm}
\includegraphics[width=4.cm, height=3.cm]{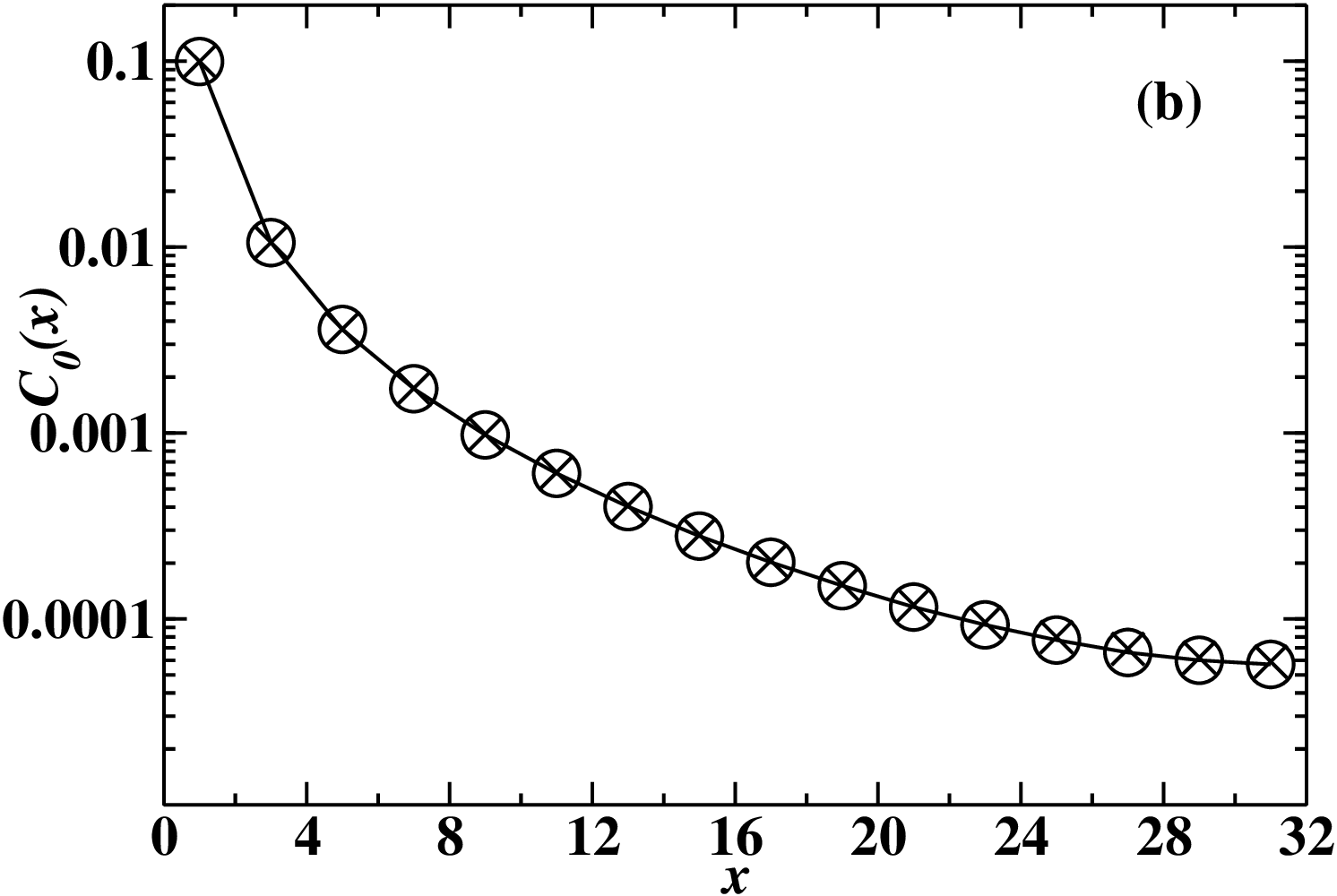}
\end{array}$
$\begin{array}{c@{\hspace{0.5in}}c}
         \multicolumn{1}{l} {}\\ [-0.23cm]
\includegraphics[width=4.cm, height=3.cm]{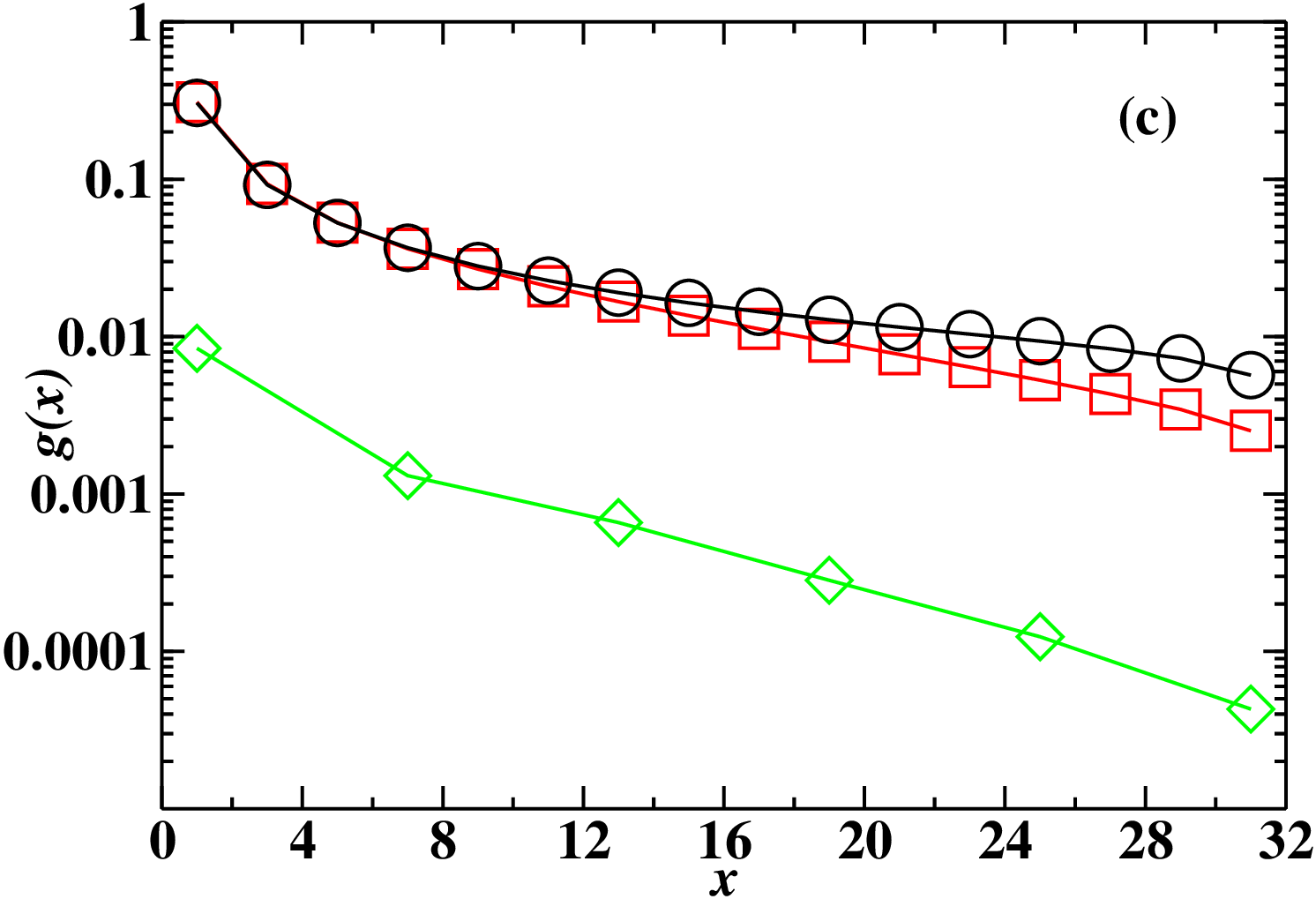}
\hspace{0.5cm}
\includegraphics[width=4.cm, height=3.cm]{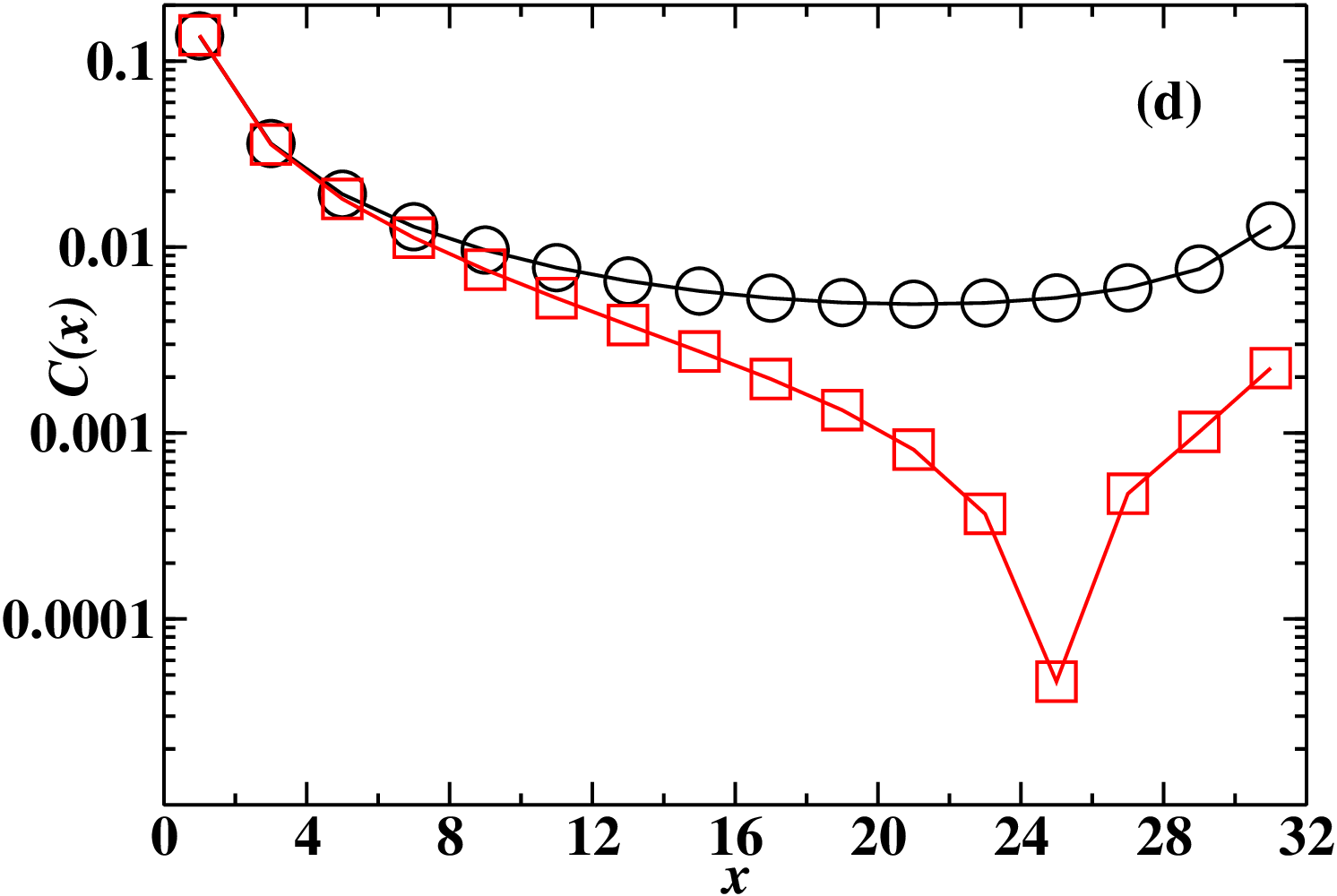}
\end{array}$
\end{center}
\caption{(a),(b), respectively $g_0(x)$, $C_0(x)$ for a $64 \times 65$ system,
for $V=0$, $t_y=0.05$, the origin is on middle of the lattice, longitudinal direction:  
 DMRG (circles), the exact result (crosses); transverse direction (only $g_0(x)$ is shown):  
 DMRG (squares), exact result (pluses). (c),(d) respectively $g(x)$, $C(x)$ for a $63 \times 63$
 system, longitudinal direction: for $V=1.5$, $t_y=0$ (circles),  for $V=1.5$, $t_y=0.1$ (squares); 
 transverse direction (only $g(x)$ is shown): for $V=1.5$, $t_y=0.1$ (diamonds).} 
\label{gc}
\end{figure}

 In order to limit the memory load, we computed the correlation functions in the central chain along the
$x$ direction, $i_y=L_x/2+1$ (longitudinal direction), and in the central chain along the $y$ direction, $i_x=L_x/2$
(transverse direction). The local density 
 $n_0(x)$ (Fig.\ref{local}), $g_0(x)$ (Fig.\ref{gc}), and $C_0(x)$ (Fig.\ref{gc}) computed with the DMRG show a very good 
agreement with the exact result. We verified that the asymptotic behavior $x \gg 1$, $g_0(x) \propto 1/x$ and $C_0(x) \propto 1/x^2$  of the
exact result is satisfied by the DMRG. For instance in Fig.\ref{gc}(a,b), for $g_0(x)$ the largest difference $\delta g_0$ between the DMRG and the exact result is seen 
in the tranverse direction at the largest distance $x=31$ for which $\delta g_0=0.0002$. The agreement is even better 
for $C_0(x)$, $\delta C_0(x) \approx 10^{-5}$ in the direction of the chains. For both the DMRG and the exact result,
in the transverse direction, $C_0(x)$ for $x>3$ falls below $10^{-6}$ which is the error set in the diagonalization of the Hamiltonian. Hence, it was not
shown. For this reason, we will exclusively concentrate on the correlation along the chains when analyzing the interacting case.

\section{Density-density correlations}

For $t_y=0$ and $V \neq 0$, there also exists an exact solution \cite{yang}. The model is in a LL phase for $V< 2t_x$ and
in a charge density wave (CDW) phase for $V > 2t_x$. In the LL phase, the asymptotic 
form of the Green's function is $g_{1D} (x)\approx A_g/x^{1+\alpha}$, where $\alpha$ 
is the anomalous exponent. The dominant two-particle 
correlations are the density-density, $C_{1D}(x)=A_c/x^{2K}$ with $\alpha=(K+1/K)/2-1$. 
At $V=2t_x$, there is a 1D QCP. When open boundary conditions are applied, the sites 
at the ends
 generate strong Friedel oscillations. These oscillations decay very slowly from the ends and interfere with the
normal density oscillations $C_{1D}(x)$. This behavior of the 1D system can be reproduced by the DMRG with extremely high
accuracy.

When $t_y$ sets in, it is expected that either the system will be dominated
by the single particle correlation, hence the ground state is a FL, or the density correlations $C_{1D}$
would freeze yielding an ordered two-dimensional CDW state. 
We did not find any evidence of  CDW long-range order (LRO) when we start
from the disordered 1D chain.
It is important to note that the same
method was used to study coupled spin chains and found LRO 
\cite{moukouri-TSDMRG} as expected.  Let us further discuss the reliability
of this result. 
Since the DMRG is highly efficient in the interacting 1D case, the level of accuracy in 1D between the cases $V=0$ and $V \neq 0$
is comparable. Hence, studying the full Hamiltonian \ref{hamiltonian}, which is no longer exactly solvable, with the DMRG is not more 
difficult than the $V=0$ case. The DMRG results of the 2D interacting case will be as good as those of the $V=0$ case, 
provided that $\Delta E$ does not decay sharply.  
When $V \neq 0$, an odd value of $L_x$ is chosen in order to not to frustrate the CDW correlations during the lattice growth \cite{caron}. This choice
also has the advantage of showing a sharp contrast for the behavior of the
Friedel
oscillations between 1D and 2D.
We find for instance that in the LL phase for $V=1.5$, $L_x=63$, 
$\Delta E$ increases from its $V=0$ value $\Delta E=0.53$ to $0.95$. This implies that for the same value of $t_y$, the accuracy would be better 
for the $V=1.5$ case than for the $V=0$ case. However, for $L_x=63$ and $V=1.5$, $\delta E$ increases to $0.11$. For this reason, 
we have to increase the transverse hopping to $t_y=0.1$ in order to effectively be in the 2D regime. But since the ratio 
$\Delta E/t_y$ remains close to its $V=0$ value, the accuracy should not change. That is, we expect the error in the interacting
Green's function $\delta g \approx 10^{-4}$ and the error in the interacting density-density correlation function
$\delta C \approx 10^{-5}$. This gives us a high degree of confidence in analyzing the properties of the 2D interacting system.

\begin{figure}
\begin{center}
$\begin{array}{c@{\hspace{0.5in}}c}
         \multicolumn{1}{l} {}\\ [-0.23cm]
\includegraphics[width=4.cm, height=3.cm]{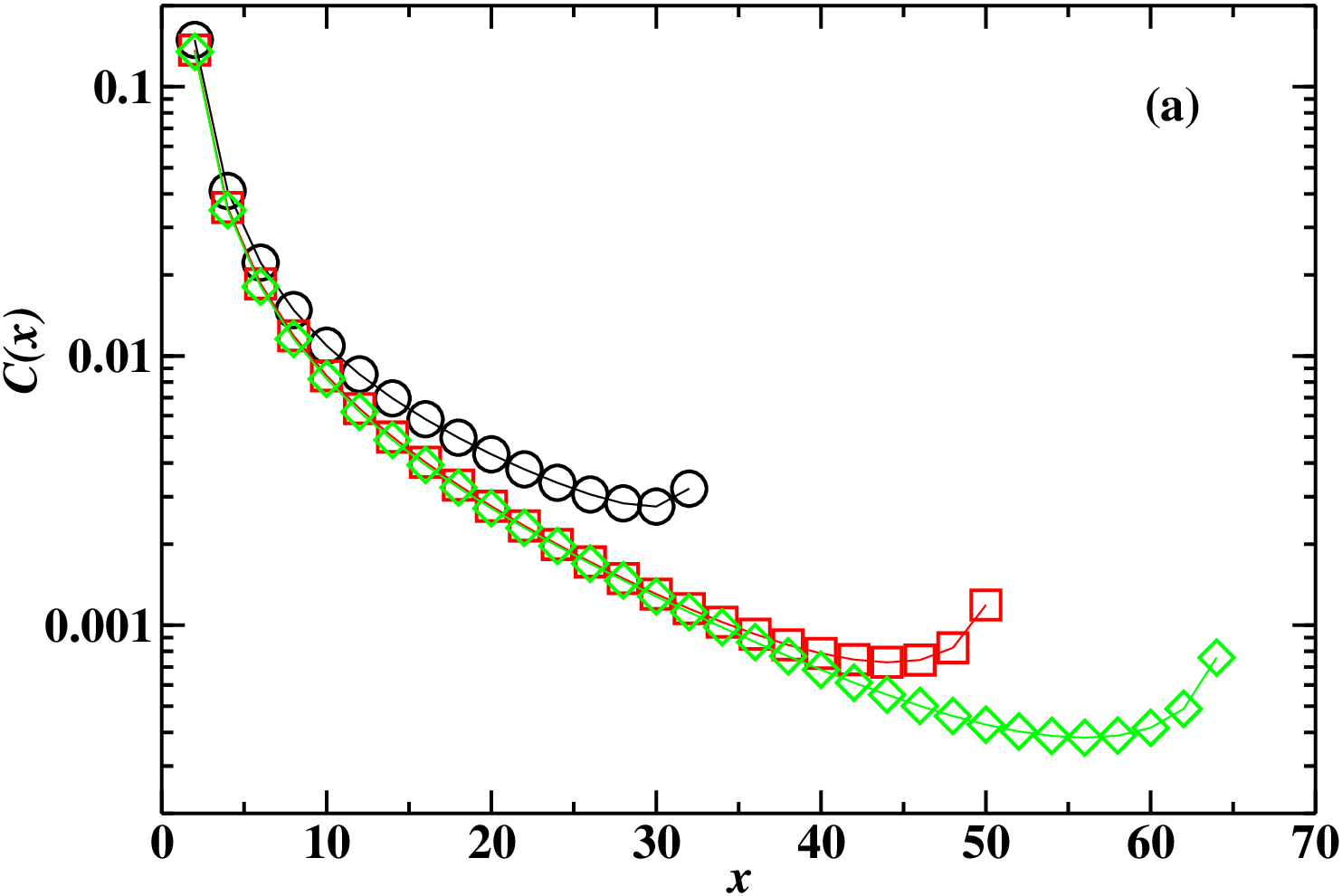}
\hspace{0.5cm}
\vspace{0.5cm}
\includegraphics[width=4.cm, height=3.cm]{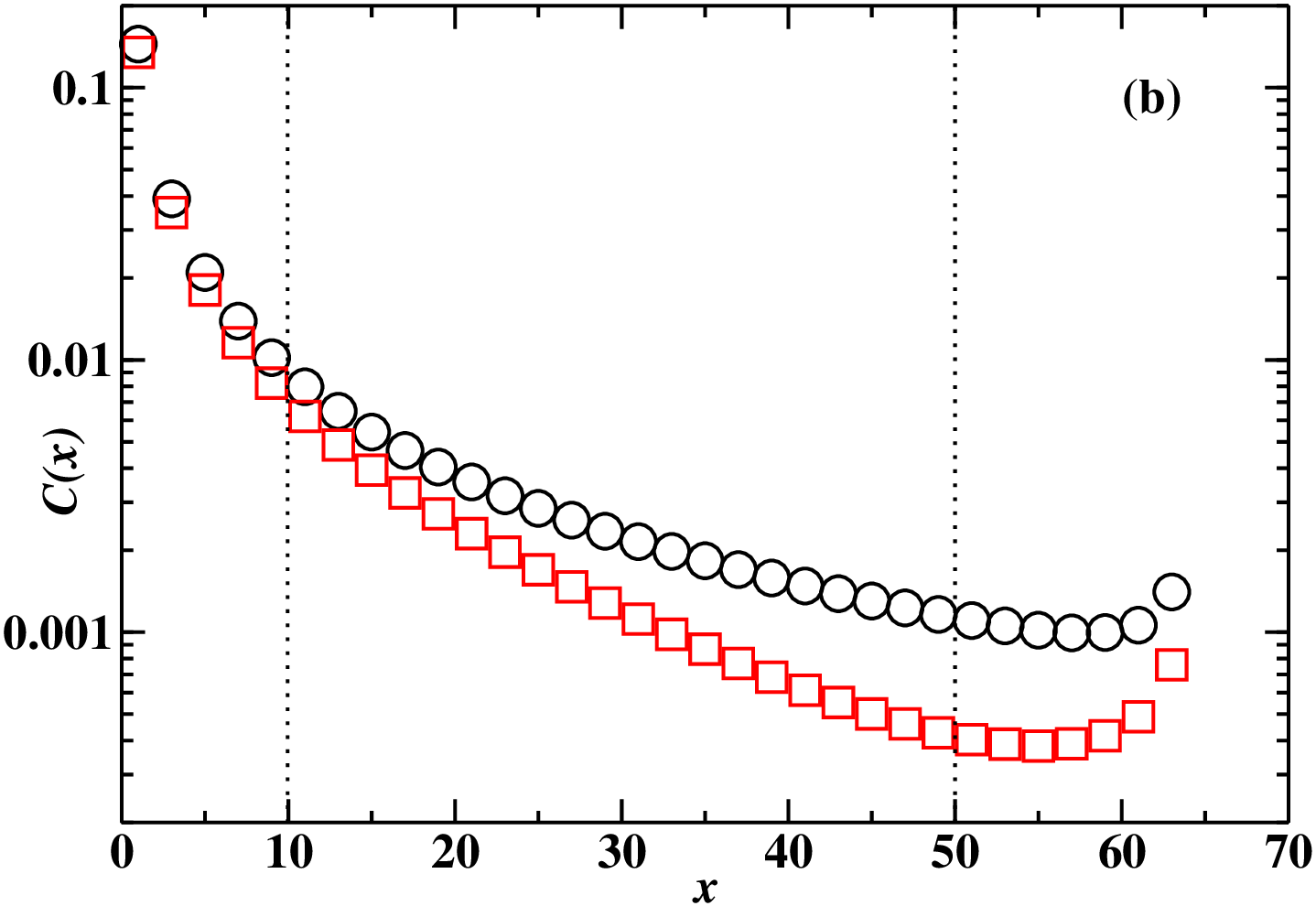}
\end{array}$
\end{center}
\caption{  $C(x)$ for $V=1.5$, $t_y=0.054$:
(a) $66 \times 67$ (circles), $102 \times 103$ (squares), and $130 \times 131$ 
 (diamonds) systems;
 (b) $L=102$ (1D, circles) and $102 \times 103$ (squares);  
the vertical dotted lines show the limit of the  
retained data used for the extrapolations.}  
\label{cc0}
\end{figure}

In the presence of $t_y$, the departure from 1D behavior which is characterized by strong oscillations of $n(x)$ 
can be seen in Fig.\ref{local}. In the 1D systems these oscillations are present even in the bulk. For
the 2D system, they vanish in the bulk, the density becomes uniform as in the case $V=0$. This is an
indication that the dramatic departure from the free electron gas seen in the interacting 1D chain is strongly
reduced by $t_y$. This is consistent with the relevance of $t_y$ or the irrelevance of $V$ (in 2D) found in 
perturbative RG. However, a crossover from a LL to a FL would be less apparent in $g(x)$ as seen in
Fig.\ref{gc}(c). This is because
 in 1D, $\alpha$ varies only from $0$ when $V=0$ to $0.25$ at the QCP $V=2t_x$.
At the same time the exponent  $K=\pi/2arccos(-V/2t_x)$, varies from $1$ to $0.5$. 
It is thus more favorable to use $C_{2D}(x)$ to analyze the crossover.
  $C_{2D}(x)$, shown in Fig.\ref{gc}(d),  has a  faster 
decay than in 1D. Unfortunately,  the actual asymptotic behavior of $C_{2D}(x)$ is masked by the presence of the remnant of the Friedel oscillations at the ends. Unlike the 1D situation, in 2D they are $\pi$-dephased with $C_{2D}(x)$.
 Hence in order to access the asymptotic behavior in 2D, we reverted to even $L_x$ with
odd $N_{ex}$, for which the Friedel oscillations are found to be less severe as
shown in Fig.\ref{cc0}.

\begin{figure}
\begin{center}
\includegraphics[width=5.cm, height=4.cm]{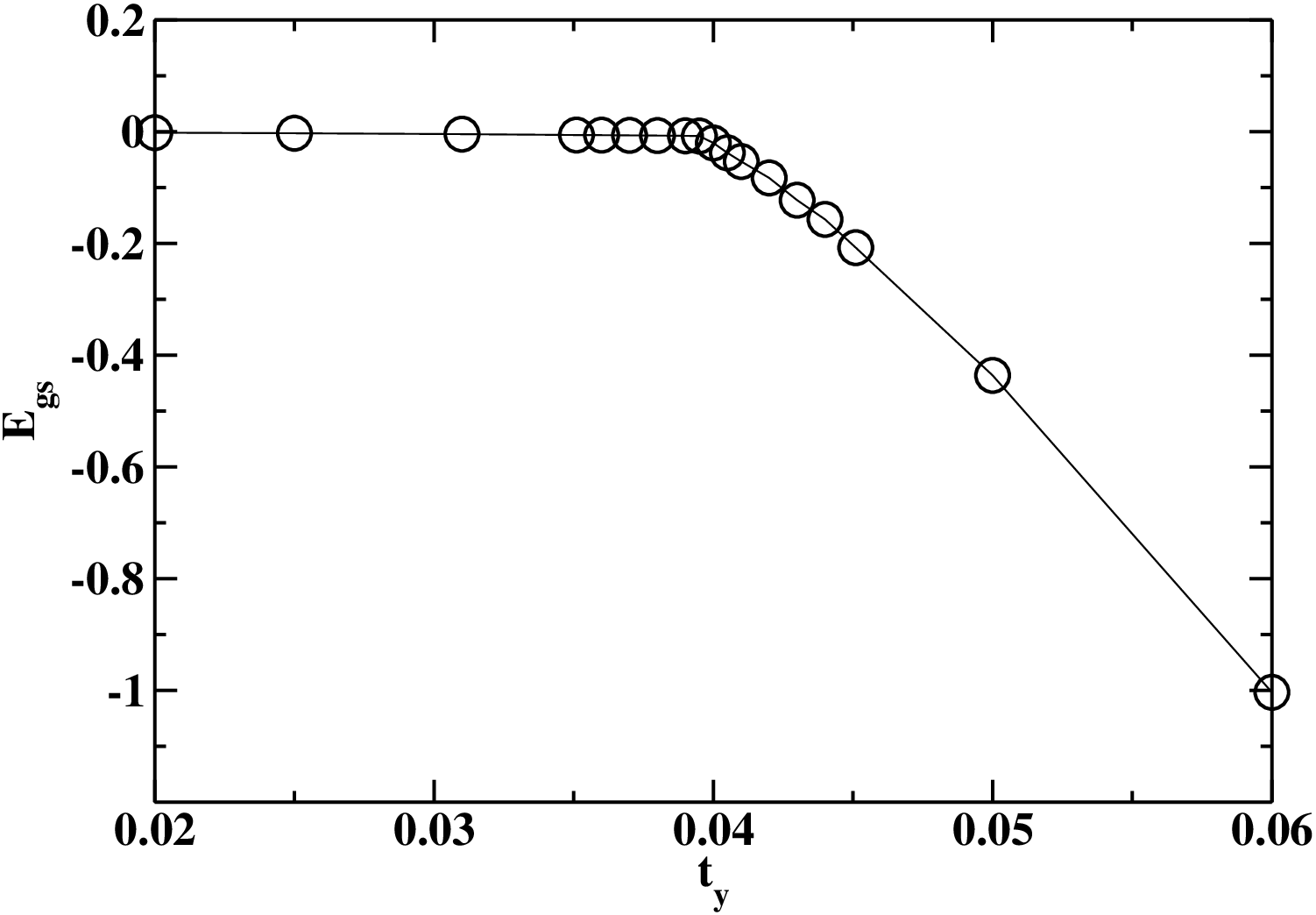}
\end{center}
\caption{  Ground state energy $E_{gs}$ with respect to the energy  
of disconnected
chains for $V=0.75$ as function of $t_y$ for
$64 \times 65$ system.}
\label{egs}
\end{figure}

\begin{figure}[b]
\begin{center}
$\begin{array}{c@{\hspace{0.5in}}c}
         \multicolumn{1}{l} {}\\ [-0.23cm]
\includegraphics[width=4.cm, height=3.cm]{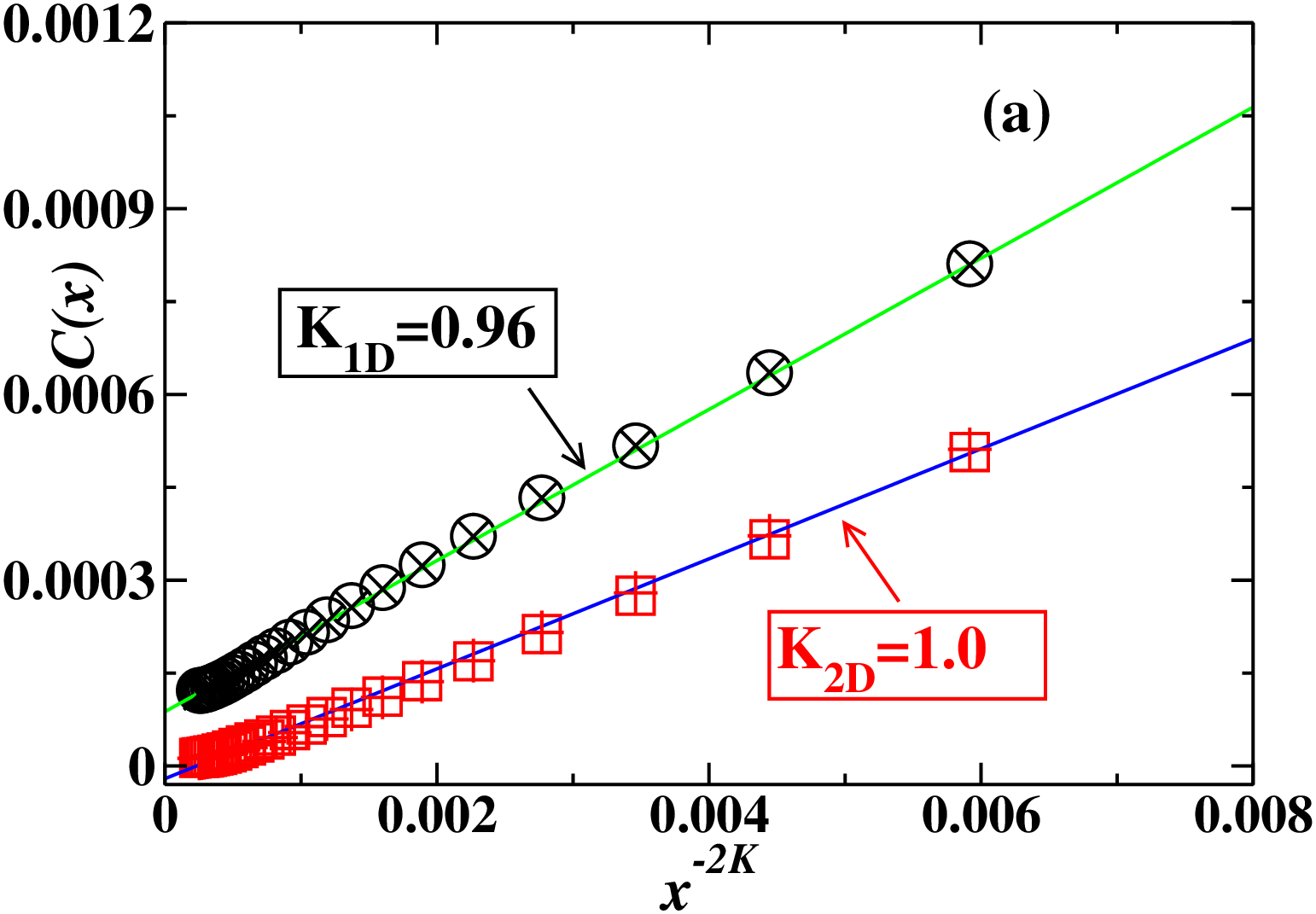}
\hspace{0.5cm}
\vspace{0.5cm}
\includegraphics[width=4.cm, height=3.cm]{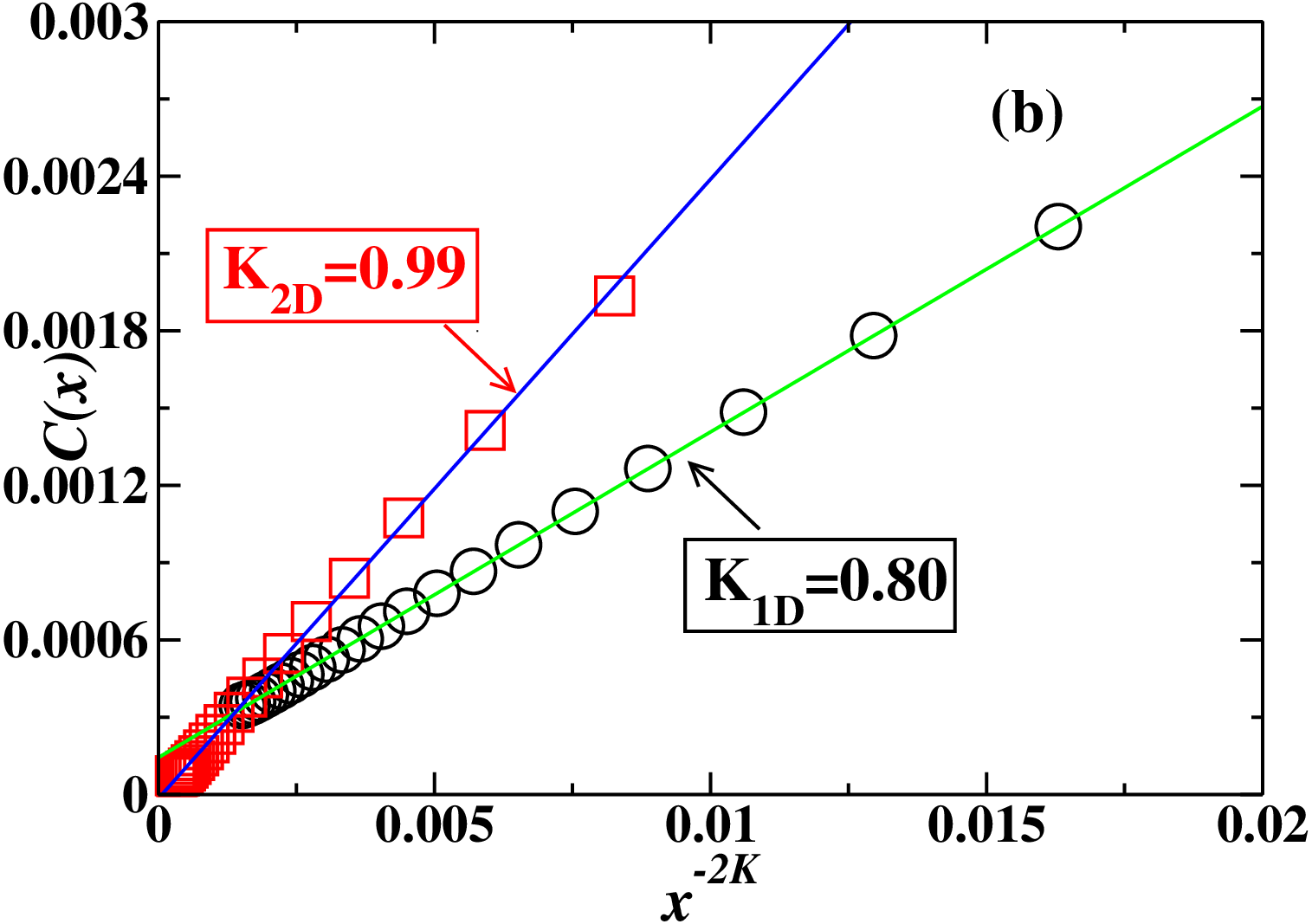}
\end{array}$
$\begin{array}{c@{\hspace{0.5in}}c}
         \multicolumn{1}{l} {}\\ [-0.23cm]
\includegraphics[width=4.cm, height=3.cm]{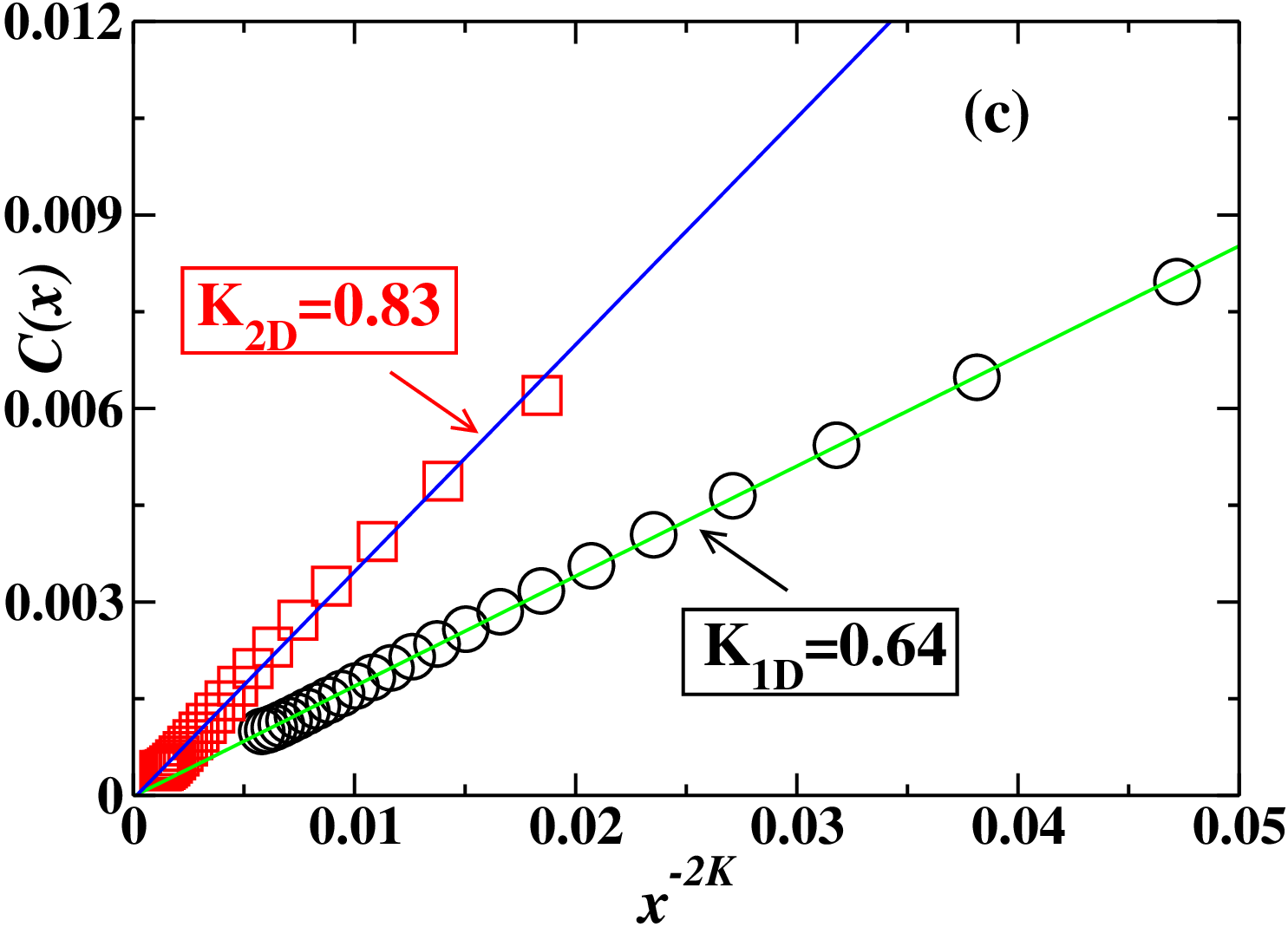}
\hspace{0.5cm}
\includegraphics[width=4.cm, height=3.cm]{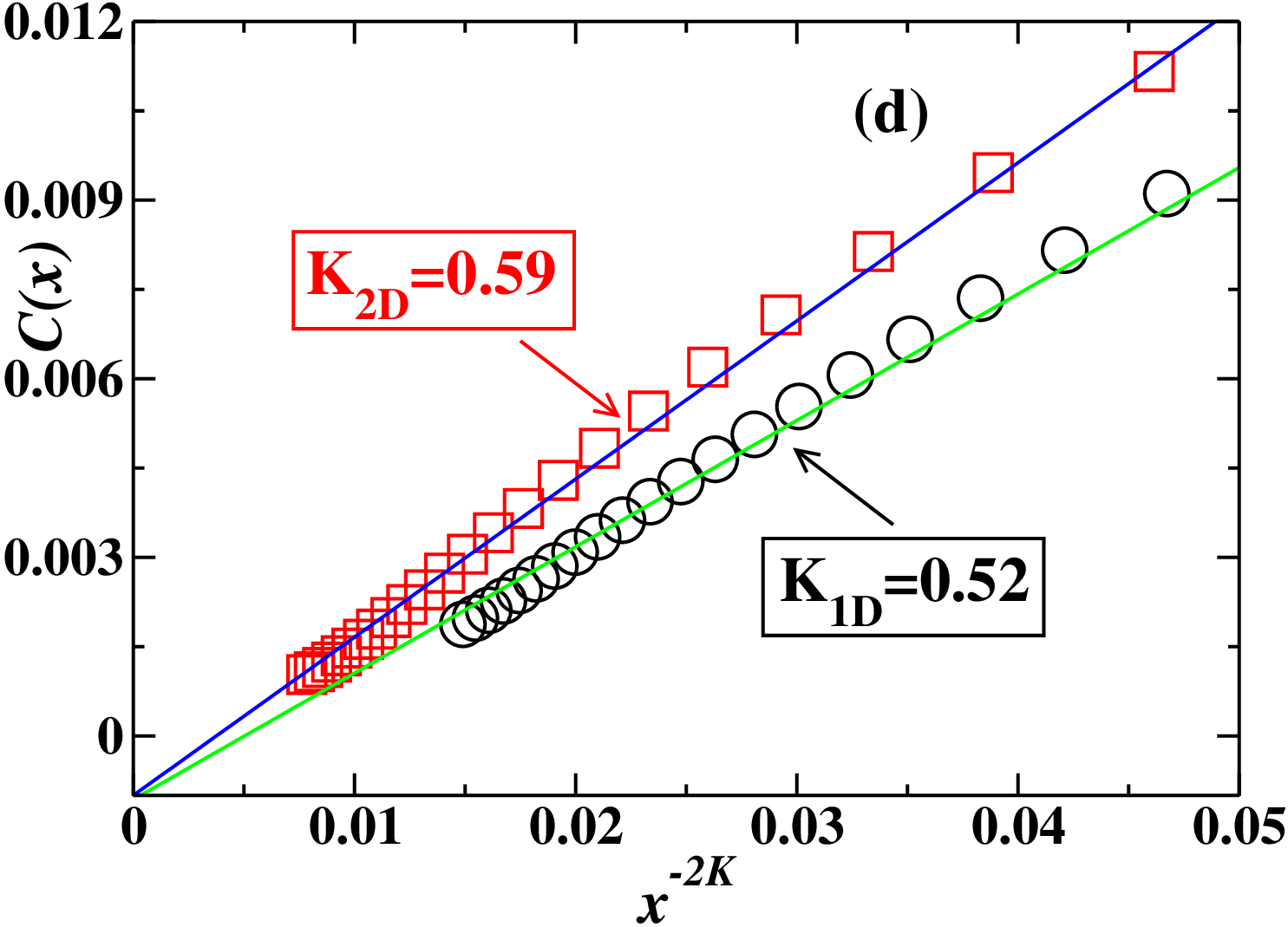}
\end{array}$
\end{center}
\caption{ Linear fit of $C(x)$ for a $130 \times 131$ systems:
in all cases 1D  DMRG (circles), exact (crosses), 2D DMRG (squares), exact (pluses); 
(a) $V=0$, $t_y=0.025$,  
(b) $V=0.75$, $t_y=0.038$; (c) $V=1.5$, $t_y=0.054$; (d) $V=2$, $t_y=0.073$.
These values of $t_y$ are chosen so that ${\tilde t_y} \approx 1$.} 
\label{cc}
\end{figure}

Once in the 2D regime, in order to extract reliable correlation exponents,
it is important to see how  finite size effects affect the decay of
correlation functions. In Fig.\ref{cc0}(a), we show $C_{2D}(x)$ for
$66 \times 67$, $102 \times 103$, and $130 \times 131$ systems for
$t_y=0.054$. $\delta E$ in these systems is respectively $0.1273$,
$0.0675$, and $0.0531$. It can be drawn from the behavior of the
$102 \times 103$ and $130 \times 131$ systems that in the regime
$t_y \agt \delta E$, aside from edges effects, finite size 
effects do not significantly affect the decay of the correlation
functions in the 2D regime. We thus believe that the exponents of
$C_{2D}(x)$ that we obtained are very close to their value in the
thermodynamic limit. 
Even if edge effects are less dramatic when an odd $N_{ex}$ is
chosen in an even $L_{x}$ system, they nevertheless strongly affect the
extraction of the correlation exponent. In Fig.\ref{cc0}(b) we show
the range of the data used for the extraction of $K$. We arbitrary
set $x \agt 10$ from the origin and from the edge. For this choice
it can be seen in Fig.\ref{cc0}(b) that we are far enough from the
upturn of $C_{2D}(x)$ caused by the edge. 

\begin{figure}
\begin{center}
\includegraphics[width=5.cm, height=4.cm]{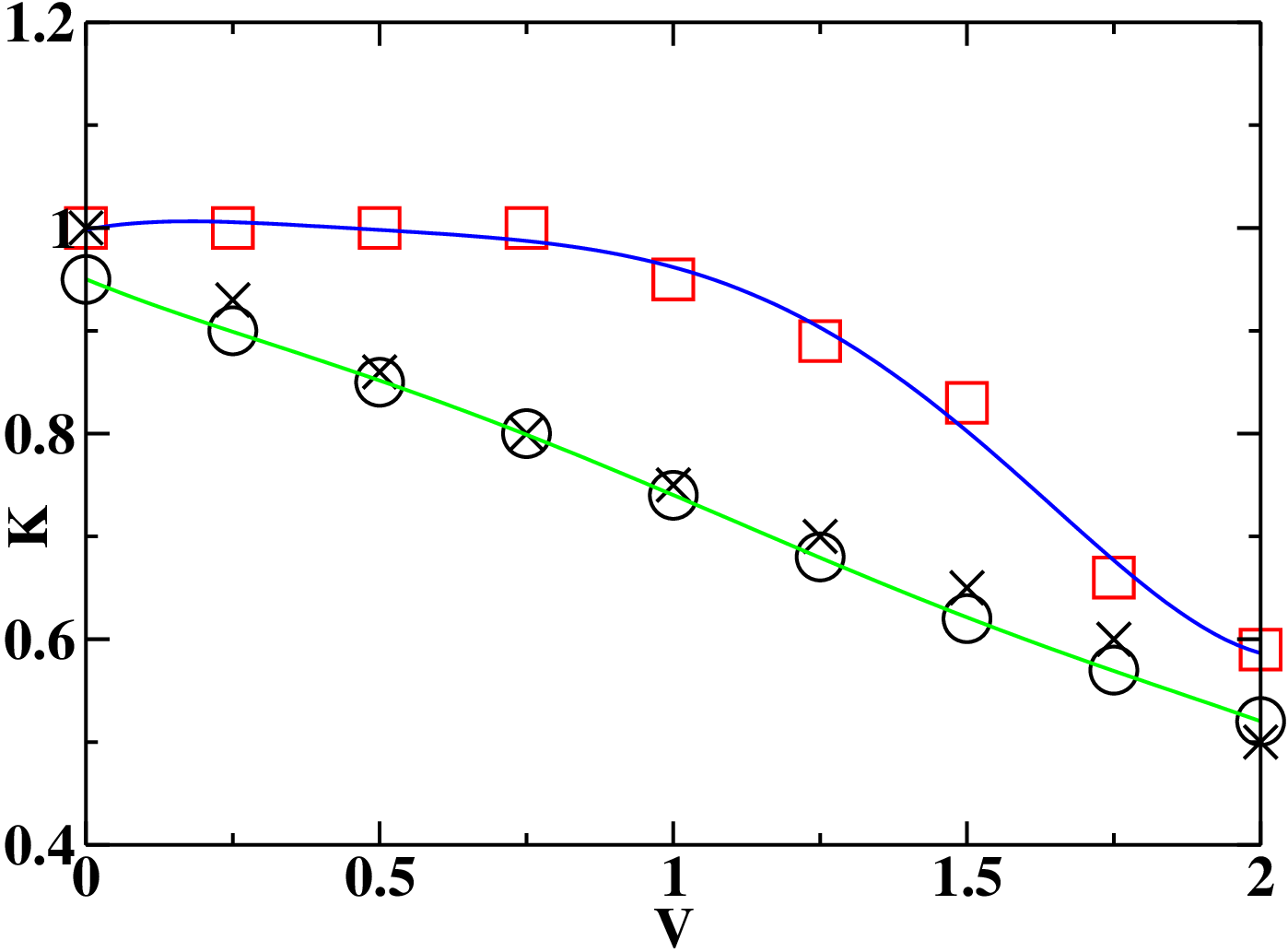}
\end{center}
\caption{ K as function of V: 1D DMRG (circles), 1D exact (crosses), 2D DMRG (squares).}
\label{corrK}
\end{figure}

Our calculations are made for small values of $t_y$, as pointed out earlier,
it is expected that when $t_y \approx \delta E$, the 2D regime is reached.
In order to make this argument more precise, we computed the ground-state 
energy $E_{gs}=E(t_y)-E(t_y=0)$ as function of $t_y$. Two regimes, shown in Fig.\ref{egs},
are observed. When 
$t_y \alt \rho \delta E$, $E_{gs}$ remains nearly equal to the energy
of disconnected chains. It would be expected that in this regime, the
system will essentially have a 1D behavior. But when $t_y \agt \rho \delta E$, 
the system
gains energy with respect to disconnected chains. This is an  
indication that the system has entered  the 2D regime. The typical value
of $\rho$ is about $0.5$ for the values of $V$ we investigated.

Since we wish to analyze the effects of $V$ on the 2D system, we must
be careful to avoid a spurious 2D to 1D crossover which is related
to finite size effects. This occurs when we increase $V$. Starting
at a relatively small value of $t_y$ and $V$ for which the condition
$\delta E \alt t_y$ is satisfied for a given size, if we increase $V$
and keep $t_y$ constant, we can reach a regime where $t_y \ll \delta E$.
Hence, we artificially enter an 1D regime. This is clearly a spurious 
effect due to the finite size of the system. In the thermodynamic
limit, $\delta E \rightarrow 0$, hence, this situation never occurs for
any finite $t_y$. This problem can be avoided by fixing the ratio
${\tilde t_y}=\frac {t_y}{\delta E}$ instead of $t_y$. This means that
we compensate for the variation of $\delta E$ induced by $V$ by increasing
$t_y$ so that we keep the system in the 2D regime.

In Fig.\ref{cc} we show the decay of $C_{1D}(x)$ and $C_{2D}(x)$ for
$V=0,~0.75,~1.5,~2$ in a $130 \times 131$ system. $t_y$ is adjusted so
that ${\tilde t_y} \approx 1$. Therefore, we are in the 2D regime
of the model. It is important to stress that, if in that case a 1D
like behavior is observed, this would be a genuine thermodynamic
behavior of the system induced by the interactions. Fig.\ref{cc}(a) shows 
that the
$V=0$ case in 1D and 2D are consistent with $1/x^2$ decay. In Fig.\ref{cc}(b),
it can be seen that the 1D data deviate from the $1/x^2$ decay. 
 On the other hand, 
the 2D behavior remains similar to the $V=0$ case. This is 
 consistent with the LL nature of the 1D system and predictions of a
FL ground state in 2D for mild interactions. When $V$ is further increased
we find in Fig.\ref{cc}(c),(d) that the 2D results deviate from the FL
behavior as well. Since there is no evidence of CDW LRO, this suggests
the existence of an unconventional metallic state in 2D.  A non-linear fit
to these data yielded $K$ which is displayed in Fig.\ref{corrK}. 
Non-linear fit are known to yield many different solutions depending
on the starting point. To avoid this problem, we first computed the
1D exponents by fixing the search range in the interval $[0,1]$. The
computed DMRG exponent shown in Fig.(\ref{corrK}) was generally in very good
agreement with the exact result. Surprisingly, the larger discrepancy was
observed at $V=0$. These 1D exponents were later used as the input in the
2D search. 
 The result shows a strong renormalization of
$K$ from  its LL value towards its FL value for $V \alt t_x$. Then,
it enters a non-FL regime with $K < 1$ when $t_x \alt V \alt 2t_x$.

\begin{figure}
\begin{center}
\includegraphics[width=5.cm, height=4.cm]{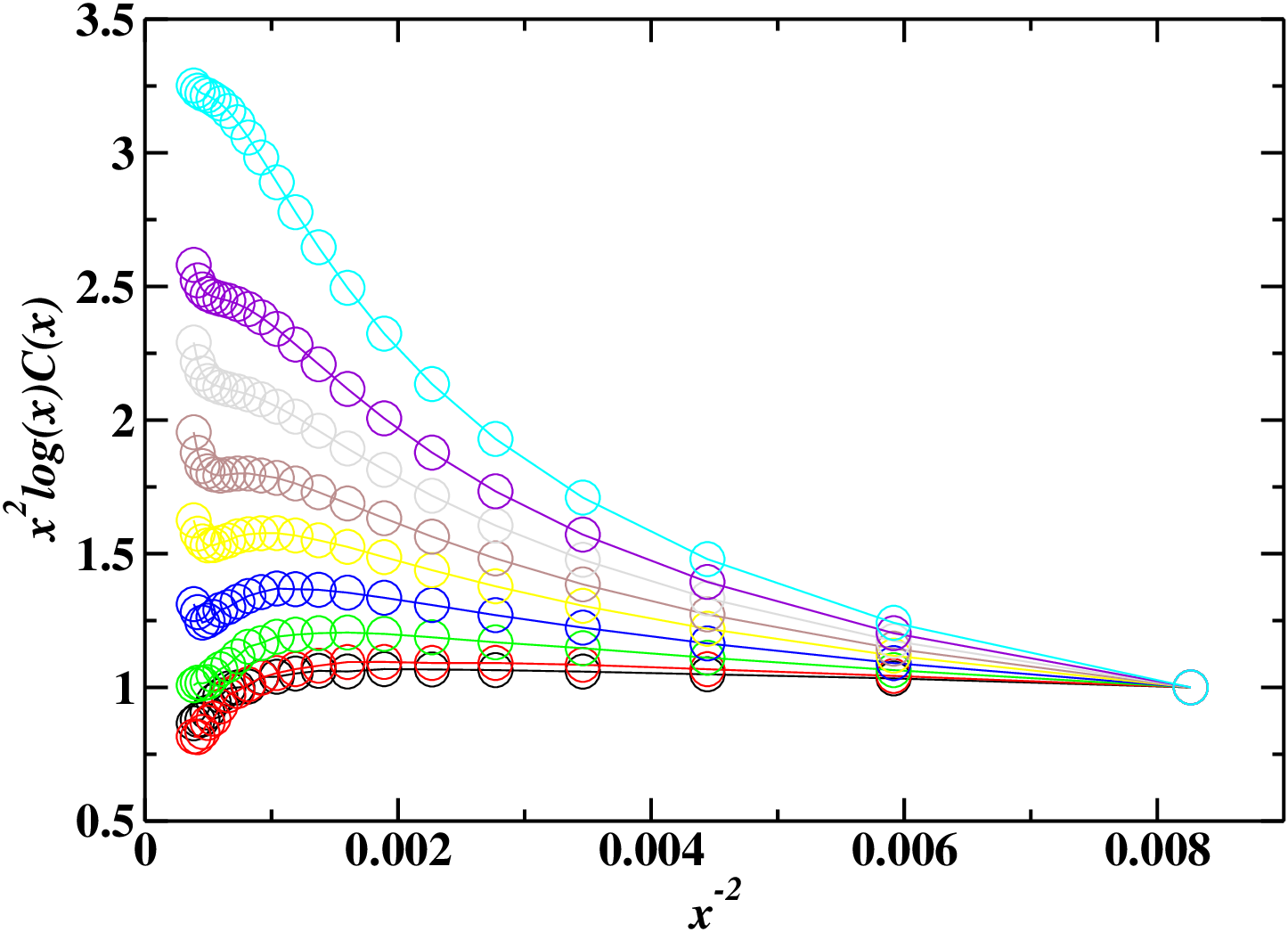}
\end{center}
\caption{ ${\tilde C}(x)=x^2log(x)C(x)$ relative to ${\tilde C}(x=11)$ 
as function of 
$x^{-2}$ for $V=0,0.25,0.5,0.75,1,
1.25,1.5,1.75,2$ (from bottom to top); 
${\tilde t_y} \approx 1$ in all cases; $t_y=0.025,0.03,0.035,0.038,0.042,
0.048,0.054,0.067,0.073$ respectively from bottom to top.}
\label{densall}
\end{figure}

In Fig.(\ref{densall}) in order to avoid the uncertainty related to
the fit, the data on $C(x)$ were directly analyzed
by studying ${\tilde C}(x)=x^2log(x)C(x)$, we added a factor $log(x)$ to the
FL  $x^{-2}$ decay to avoid a maximum that occurs in
$x^2C(x)$ at large $x$ and $V > 1.25$. We believe that this maximum
is due to logarithmic corrections in the vicinity of the QCP. The factor
$log(x)$ did not qualitatively modify the behavior of ${\tilde C}(x)$
when $V < 1$.
The plot of $\frac {{\tilde C}(x)} {{\tilde C}(11)}$ for
values of ${\tilde t_y} \approx 1$,
shows an evolution from $V=0$ to $V=2$. In the small $V$ regime, 
${\tilde C}(x)$ is nearly flat. This is consistent with the FL physics.
There is a downturn at large $x$ which is probably due to the influence
of the edges. In the vicinity of the QCP ${\tilde C}(x)$ increases steadyly
with increasing $x$. This clearly proves that $K$ is smaller than its FL
value in this regime. It should be noted that the smallnest of $t_y$ implies
that the 2D QCP remains very close to its 1D counterpart. Thus even if
$V=2$ is not exactly at the 2D QCP, it lies very close to it.
${\tilde C}(x)$ evolves between these two limits as $V$ increases. This
shows that the strong $V$ regime clearly departs from the FL picture.

\section{Low frequency spectrum}

\begin{figure}
\begin{center}
\includegraphics[width=5.cm, height=4.cm]{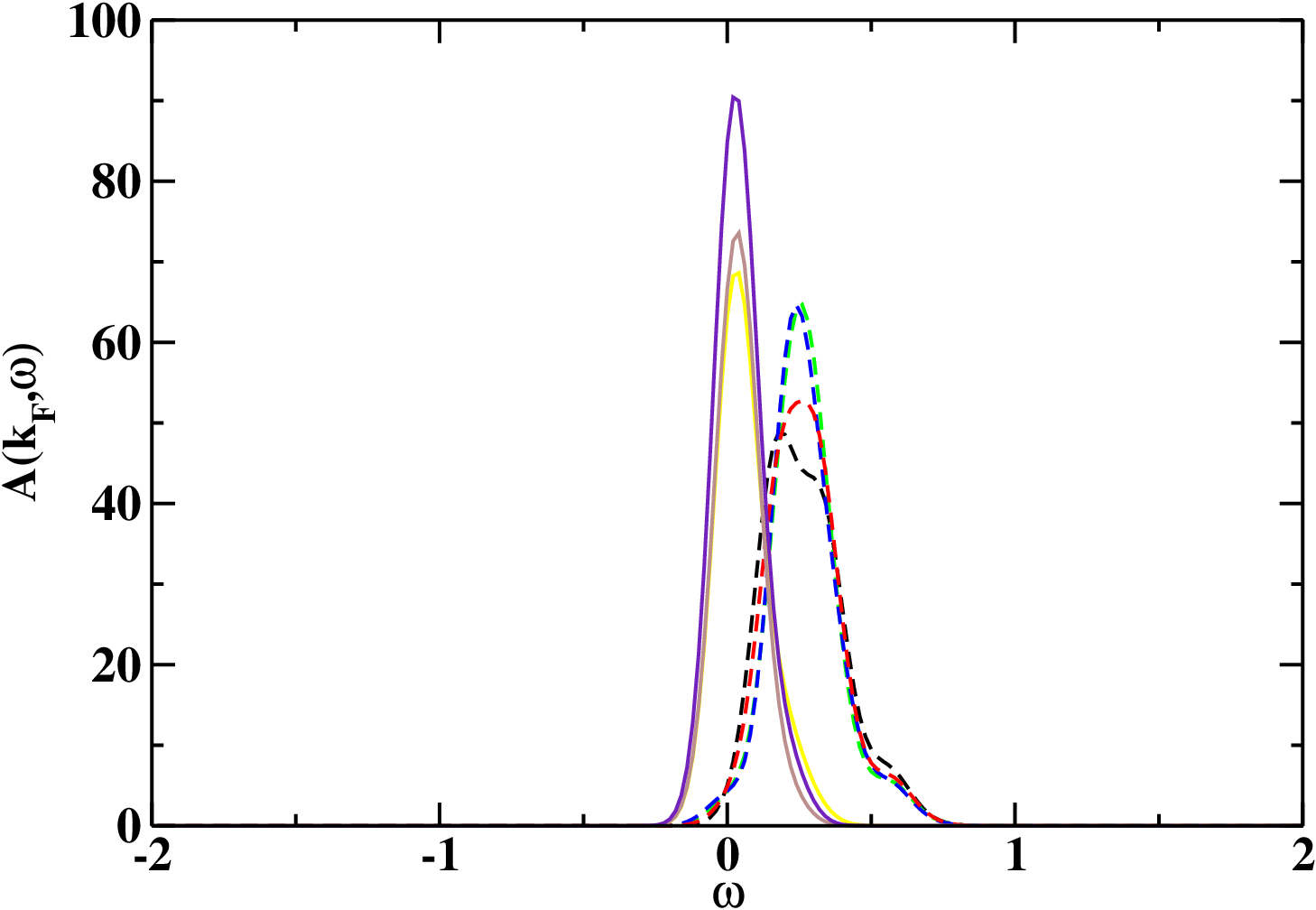}
\end{center}
\caption{ Spectral Weight function $A(k_F,\omega)$  for $64 \times 64$ systems
respectively for increasing height
for $V=0.75$: $t_y=0.005$ , $t_y=0.02$, $t_y=0.03$, and
$t_y=0.0395$ (dotted lines), $t_y=0.04$, $t_y=0.05$, and 
$t_y=0.06$ (full lines).}
\label{akwv075}
\end{figure}

A more direct information on the presence or lack of quasi-particles is given by the 
spectral weight
function, $A({\bf k},\omega)=\sum_n \langle \Psi_n| a_{\bf k}^{\dagger}|\Psi_0 \rangle \delta(\omega-\zeta_{n0})$, where $\Psi_0$ is the ground-state
wave-function,
$\Psi_n$ are the excited state wave functions with $N_e+1$ electrons, and 
$\zeta_{n0}$
are the excitation energies between the levels $0$ and $n$. The scope of finite
frequency study will necessarily be limited to very low
frequencies. The multiple RG steps in 1D and 2D have truncated out 
most of the Hilbert space of the system. We are left with a very tiny
fraction of the total number of eigenfunctions and eigenvalues. The
essential goal of this section is to show that $A({\bf k},\omega)$ near $\omega=0$
is consistent with our conclusions on $C_{2D}(x)$.
 
The DMRG can yield the low $\omega$ behavior of $A({\bf k},\omega)$ by targeting lowest states of sectors with $N_e$ and $N_e+1$ electrons. The low energy spectrum is then obtained 
by diagonalizing the reduced superblock of size $ms_2 \times ms_2$ made of the two external
blocks.  
As for the density-density correlation, we restricted ourselves to the central chain
 and used the following approximation, $A( k_x,\omega)=\sum_n \langle \Psi_n| a_{ k_x}^{\dagger}|\Psi_0 \rangle e^{-\frac{(\omega-\zeta_{n0})^2}
{\xi^2}}/\sqrt{\pi \xi^2}$, with $\xi^2 =10^{-2}$. This means that we are not
exactly at the Fermi point of the 2D systems. In presence of $t_y$, the
Fermi point along $k_y=0$ is $(k_F+\delta k_F,0)$, with $\delta k_F=Arccos(-ty/tx)-k_F$.
Since $t_y/t_x \ll 1$, $\delta k_F$ is very small. The 1D Fermi point
remains very close to the 2D Fermi surface.

\begin{figure}
\begin{center}
$\begin{array}{c@{\hspace{0.5in}}c}
         \multicolumn{1}{l} {}\\ [-0.23cm]
\includegraphics[width=4.cm, height=3.cm]{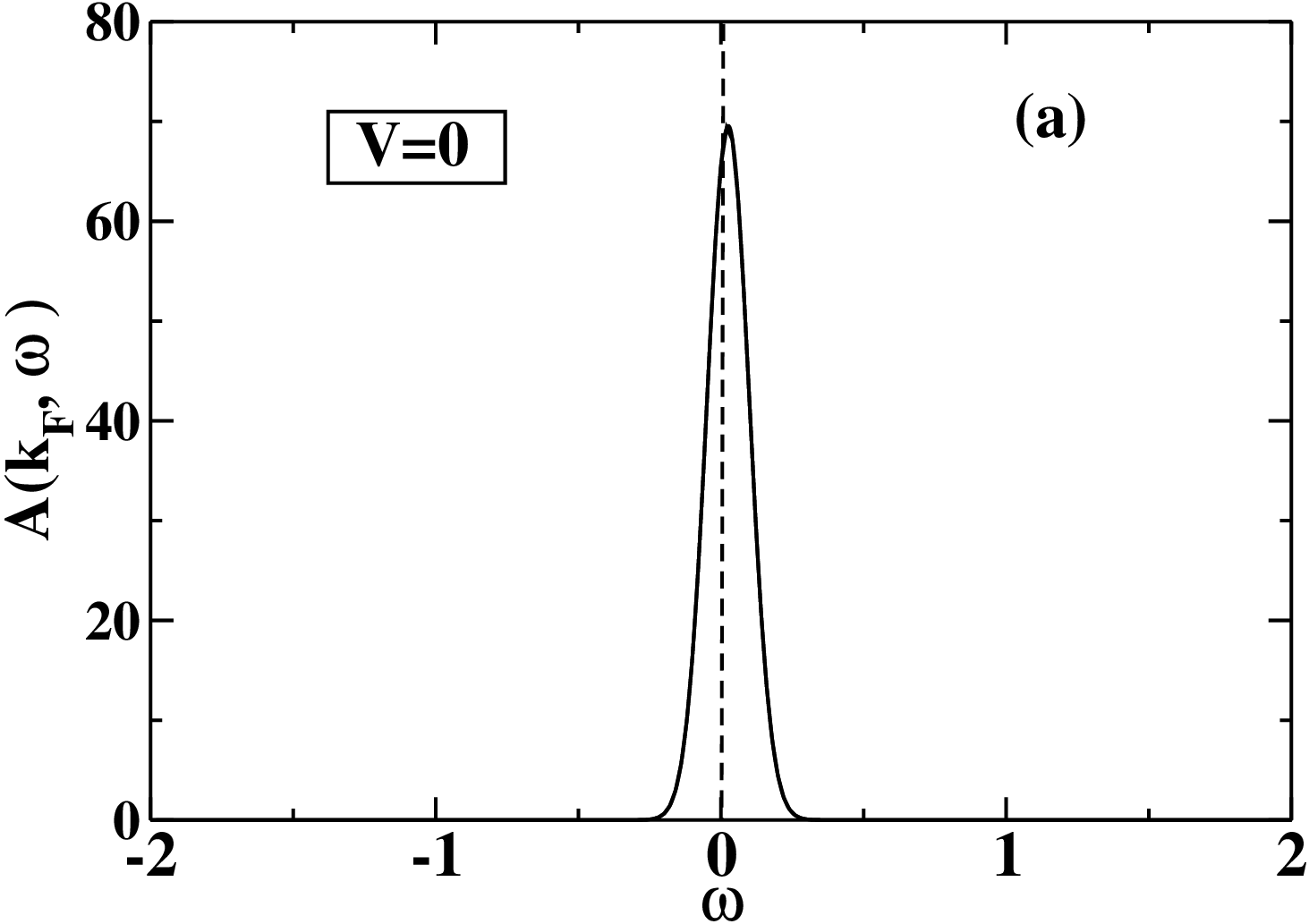}
\hspace{0.5cm}
\vspace{0.5cm}
\includegraphics[width=4.cm, height=3.cm]{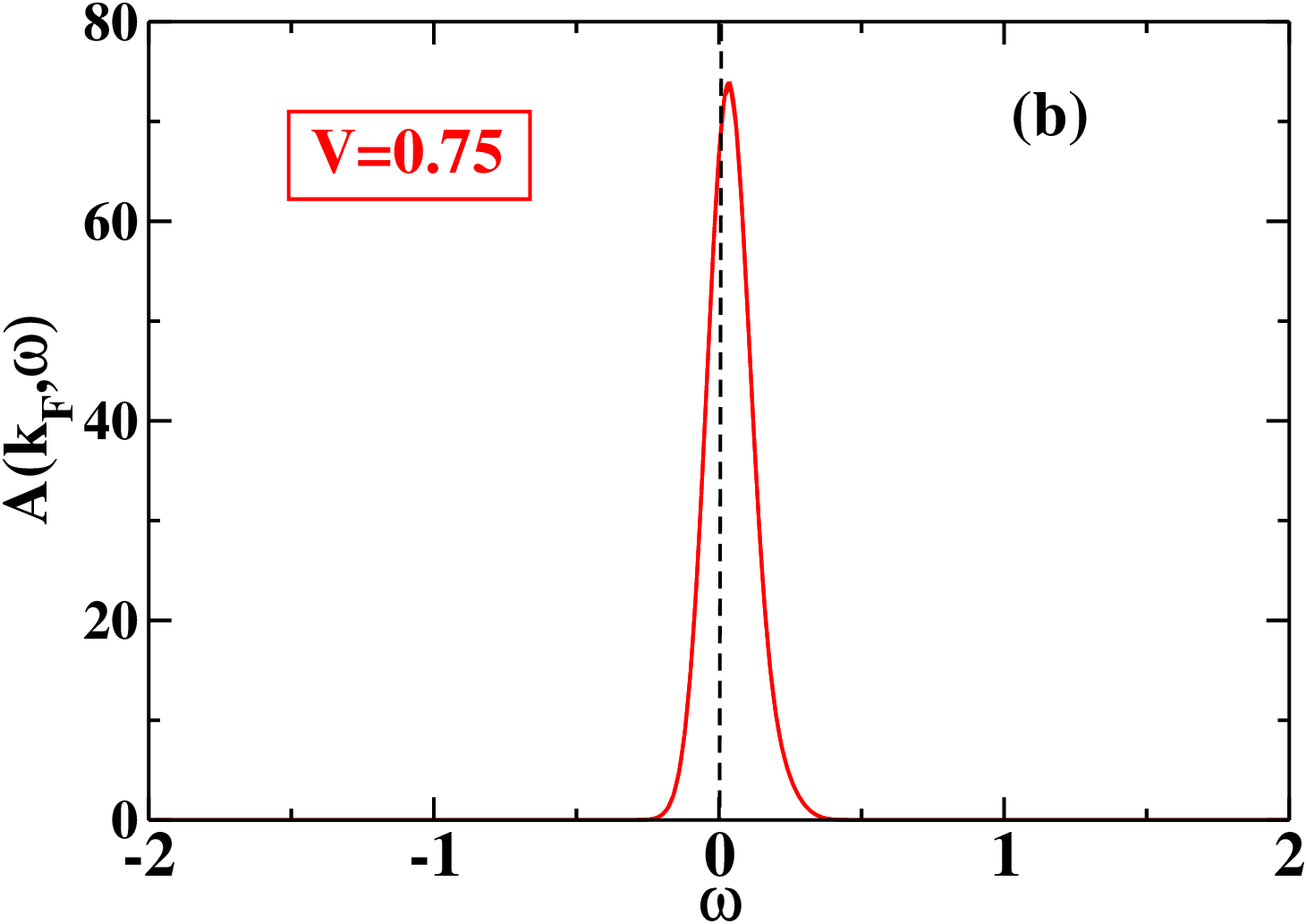}
\end{array}$
$\begin{array}{c@{\hspace{0.5in}}c}
         \multicolumn{1}{l} {}\\ [-0.23cm]
\includegraphics[width=4.cm, height=3.cm]{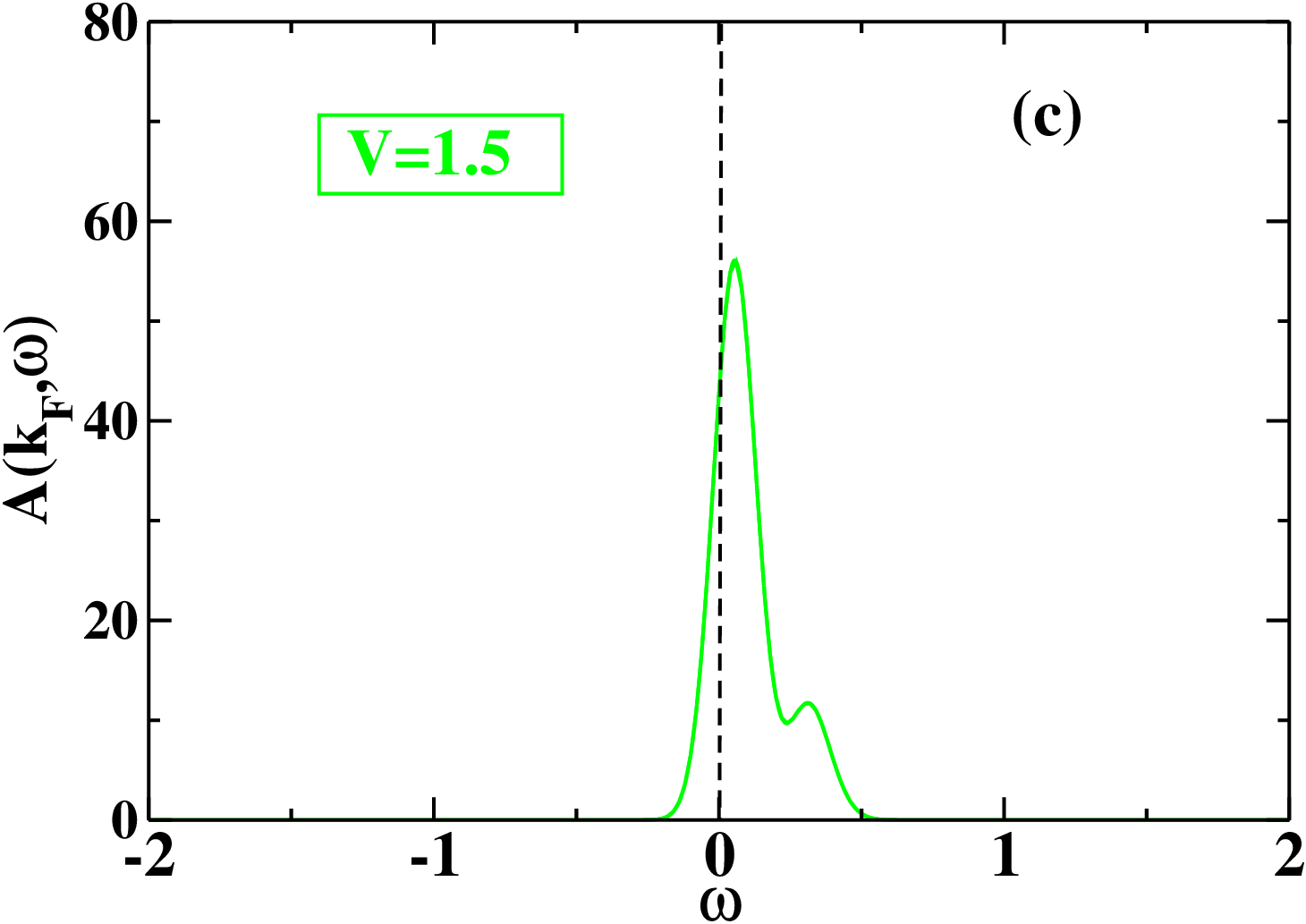}
\hspace{0.5cm}
\includegraphics[width=4.cm, height=3.cm]{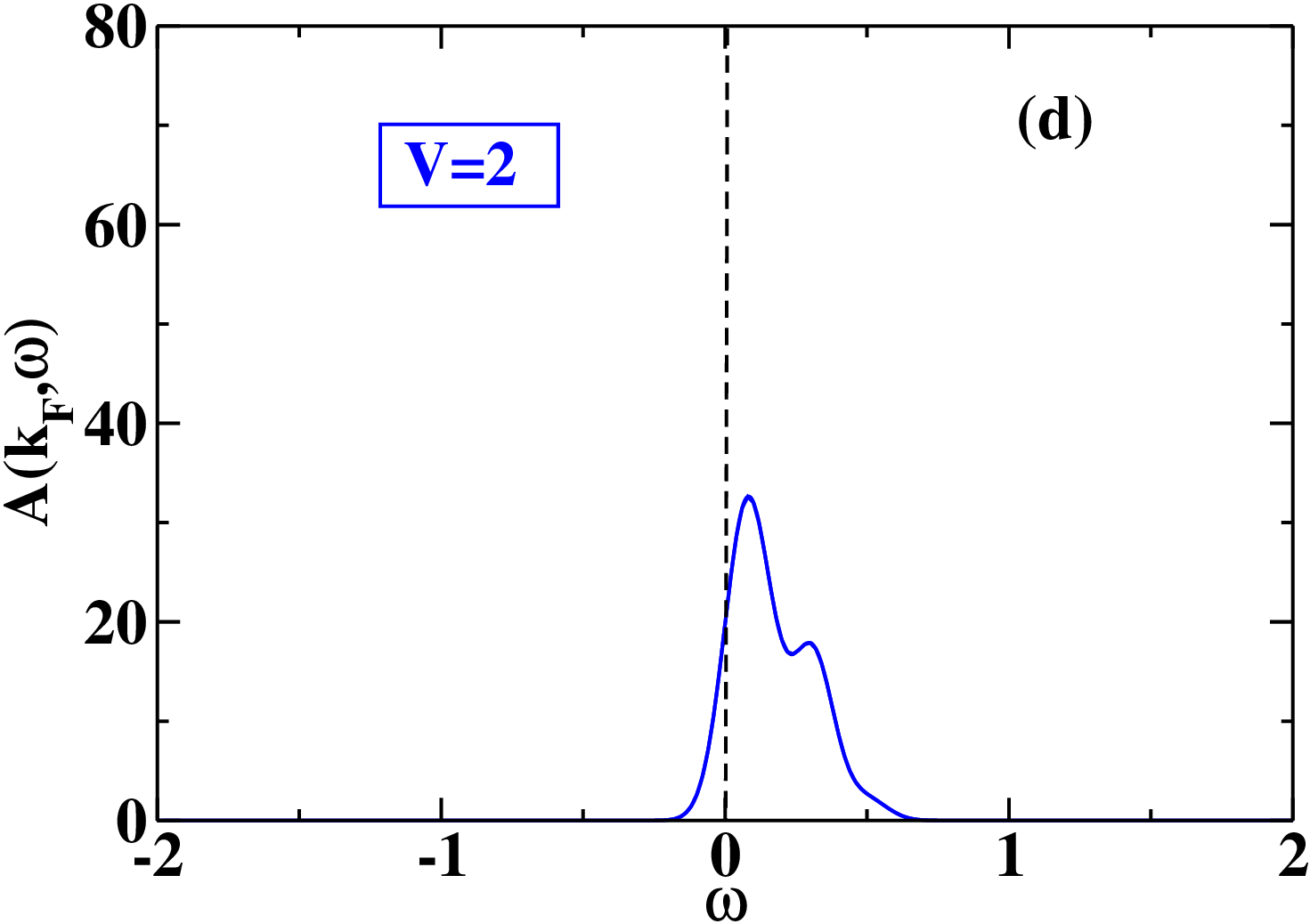}
\end{array}$
\end{center}
\caption{Spectral Weight function $A(k_F,\omega)$  for $64 \times 64$
 systems: (a) $V=0$, $t_y=0.05$; (b) $V=0.75$, $t_y=0.05$;
  (c) $V=1.5$, $t_y=0.075$; (d) $V=2$, $t_y=0.15$.}
\label{akw}
\end{figure}

As in the case of $C_{2D}(x)$, if $t_y \ll \rho \delta E$, $A({\bf k},\omega)$,
the system will not show the 2D behavior. This is seen in spectra displayed
in Fig.\ref{akwv075} for $V=0.75$ and $L_x \times L_y=64 \times 65$. When
$t_y \alt 0.04$ (this is the point where a cusp is seen in $E_{gs}$), there is a pseudogap in $A({\bf k},\omega)$. This pseudogap
is a 1D finite size effect. And as soon as the $t_y \agt 0.04$, i.e the
system enters the 2D regime following the criterion on $E_{gs}$, a
quasi-particle peak appears in $A({\bf k},\omega)$. Hence, to some
extent if L is large enough, increasing $t_y$ is
equivalent to increasing the size of the system. This transition is very
sharp. This is consistent the cusp observed in $E_{gs}$. In the 2D
regimes the peak becomes sharper when $t_y$ is increased. This was
observed for $0.04 \alt t_y \alt 0.075$. When $t_y \agt 0.075$, the
peak was less sharp. This is due to the fact that for this value of
$t_y$, the condition $t_y \ll \Delta E$ was no longer satisfied, hence
we started loosing accuracy.

The spectra shown in Fig.\ref{akw} are consistent with the
conclusions drawn from $C(x)$. When $V \alt t_x$, there is a well defined quasi-particle
peak at the Fermi energy for $k_x=k_F$. But when $t_x \alt V \alt 2t_x$,
the peak shifts away from the Fermi energy; $A(k_F,\omega)$ displays a
quantum-fluctuation induced pseudogap which is a precursor of the CDW gap. 

\begin{figure}
\begin{center}
\includegraphics[width=5.cm, height=4.cm]{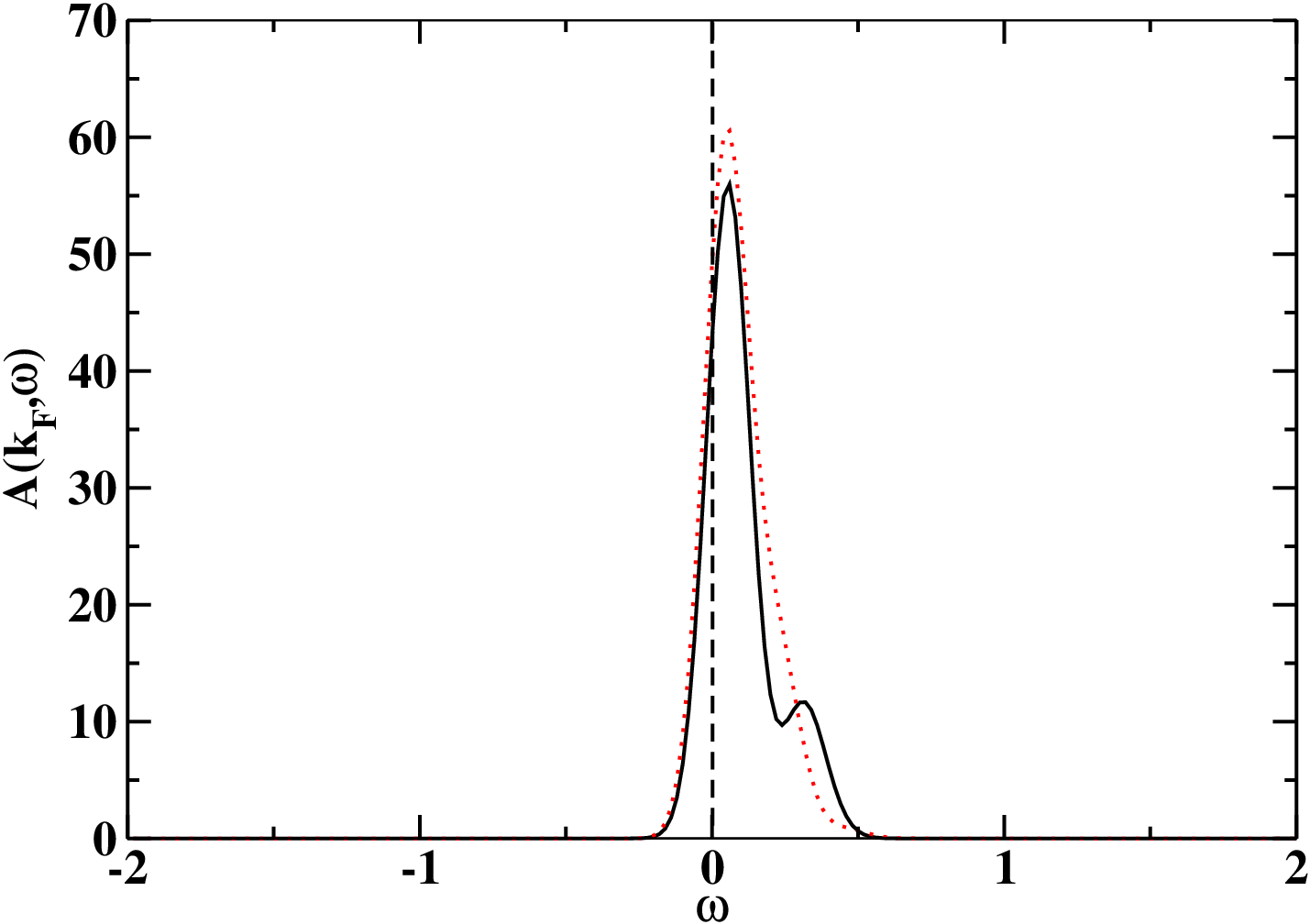}
\end{center}
\caption{ Spectral Weight function $A(k_F,\omega)$  for $64 \times 64$ systems
for $V=1.5$: $t_y=0.075$ (full line), $t_y=0.1$ (dotted line).}
\label{akwv15}
\end{figure}

The pseudogap observed between $1 \alt V \alt 2$ exists for a finite range
of $V$. This is for instance illustrated  in Fig.(\ref{akwv15}) where
$A(k_F,\omega)$ is shown for $t_y=0.075$ and $t_y=0.1$. A pseudogap
exists for both values of $t_y$. For $t_y=0.1$, there is a shift of the
spectral weigth towards lower $\omega$. This is consistent with the
possible crossover towards a FL at higher $t_y$.

\section{Discussion and conclusion}

In this work, we have shown that the two-step DMRG is a very useful tool 
for the study of
quasi-1D models. This was illustrated in a spinless fermion model.
We were able to obtain a very good accuracy on systems as large as
$L \times (L+1)=130 \times 131$. This is out of the reach of the 
conventional DMRG algorithm \cite{white}. The essential difference with
the coventional DMRG is the separation of the two energy scales of
the Hamiltonian. The difficulty that arised
in the interpretation of data was due not to accuracy but rather to
the effects of open boundary conditions. In principle
this could be avoided if periodic boundary conditions are used. 
However, we find that even periodic boundary conditions were not free of
problem. When they are applied, the ground
state is two-fold degenerate for even $N_e$. This means that
in constructing the 1D Hamiltonian, additional effort should be made to
keep tract of these multiplets. The computed low energy Hamiltonian for
a single chain is then less accurate than with open boundary conditions.

The analysis of the
correlation functions and of the low-energy spectral weight function 
revealed that 
 a FL behavior occurs for small interactions. However, when the values of the
interaction are close to the 1D quantum critical point, $V=2t_x$, the system behavior
departs from that of a FL. A conservative view would be to infer
this discrepancy to finite size effects. We did our best to disprove this
possibility by showing that the decay of $C(x)$ is essentially identical
for $102 \times 103$ and $130 \times 131$ systems for $V=1.5$. Given 
the nearly size independence of $K$ between $102 \times 103$ and $130 \times 131$, 
it is unlikely that the value $K=0.83$ at $V=1.5$ and ${\tilde t} \approx 1$
would significantly change to
reach the FL value $K=1$ in the thermodynamic limit.

 These results could be explained by the unconventional ideas raised by
Anderson and coworkers \cite{clarke,strong}. When $0 \alt V \alt 1$,
the LL would be unstable against any small $t_y$. The resulting state
is a FL. However, when $1 \alt V \alt 2$, though $t_y$ is a relevant
perturbation, there are no quasi-particles until $t_y$ exceeds a certain
treshold.

\begin{acknowledgments}
 This work was supported in part by the Israel Science
Foundation through grant no. 1524/07 and by the Ministry
of Immigrant Absorption. 
 We  thank V. Lieberman 
 for reading the manuscript. We acknowledge useful discussions with D. Orgad and
helpful correspondence with
J. Zaanen, A.-M.S. Tremblay.
\end{acknowledgments}

\end{document}